\documentclass[iop,apj,tighten]{emulateapj}
\usepackage{apjfonts} 
\usepackage{amsmath,amstext,soul}
\usepackage[breaklinks,colorlinks,citecolor=blue,linkcolor=magenta]{hyperref} 
\usepackage[all]{hypcap} 
\usepackage[normalem]{ulem}
\usepackage{color}

\shorttitle{relativistic radiation hydrodynamics code}
\shortauthors{Rivera-Paleo \& Guzm\'an}

\begin{document}

\title{{\it CAFE-R} a code that solves the special relativistic radiation hydrodynamics equations}

\author{F. J. Rivera-Paleo and F.S. Guzm\'an  }
\affil{Instituto de F\'{\i}sica y Matem\'{a}ticas, Universidad Michoacana de San Nicol\'as de Hidalgo. Morelia, Michoac\'{a}n, M\'{e}xico.}

\begin{abstract}
We present a 3D special-relativistic radiation hydrodynamics code. It uses the radiative  inversion scheme with the M1-closure relation for the radiation equations, which allows the treatment of a wide range of optical depth, temperature and opacity. The radiation field is treated in the grey-body approximation. We present the standard 1D and 2D tests that include both optically thin and thick scenarios, as well as  hydrodynamical and radiation pressure dominated configurations. As an application in 3D, we show the evolution of a  jet driven by radiation-hydrodynamics with a helical perturbation. The code is expected to allow the exploration of scenarios in high-energy astrophysics where the radiation is important, like sources of GRBs.
\end{abstract}

\maketitle


\section{Introduction}
\label{sec:Introduction}

The radiation is ubiquitous in all high-energy astrophysical scenarios and in many of them, radiative transference  has large effects in the dynamics of the system, specially in scenarios where the radiation is strongly coupled with the matter. Examples of scenarios like these include core-collapse supernovae and supernova shock breakout \citep{Shibata_et_al_2011,Kuroda_et_al_2012,Suzuki_et_al_2016,Kuroda_et_al_2016,Roberts_et_al_2016,Obergaulinger_et_al_2018,Skinner_et_al_2018}, jets from active galactic nuclei (AGN), microquasars, tidal disruption events (TDEs) and gamma ray bursts (GRBs) \citep{Aloy_Rezzolla_2006,Cuesta_et_al_2015,Cuesta_et_al_2015b,Rivera_Guzman_2016,DeColle_et_al_2017,Rivera_Guzman_2018,Aloy_et_al_2018}, accretion of material onto black holes \citep{Zanotti_et_al_2011,Fragile_et_al_2012,Sadowski_et_al_2013b,Parfrey_Tchekhovsko_2017,Fernandez_et_al_2018,Liska_et_al_2018,Dai_et_al_2018}, as well as merger and post-merger of compact objects \citep{Foucart_et_al_2015,Sekiguchi_et_al_2016,Fujibayashi_et_al_2017,Kyutoku_et_al_2018}. In the most complex cases, the numerical models that emulate the radiation field of these high-energy astrophysical phenomena should include the solution of the Boltzmann radiation transfer equation, for either photons or  neutrinos.

However, solving the Boltzmann equation for photons/neutrinos in 3D is a demanding challenge. There are two approximate methods that are successfully applied to solve the Boltzmann radiative equation, which describe the optically thin and/or optically thick regimes pretty well. The first one of them is the so-called Eddington approximation, in which the zeroth and  first-moment equations of the Boltzmann equation are solved and closed by the Eddington tensor, which is evaluated assuming  the radiation field is isotropic. This is a useful approach to treat systems in the diffusion limit \citep{Mihalas_Mihalas_1984}. Otherwise, in the free-streaming limit, the Eddington approximation does not work properly because the speed of signals is limited by $1/\sqrt{3}$. In the second approximate method, the zeroth and the first-moment equation are closed with a more general assumption about the Eddington tensor, considering  anisotropies in the radiation field; in this case, the Eddington tensor is obtained as a function of  zeroth and first moments and allows the radiation to propagate with the speed of light in optically thin media \citep{Levermore_1984}. This second approach is called the M1-closure. 

These two approximate methods have been widely used to couple radiation with matter. For instance, in the non-relativistic radiation hydrodynamics limit \citep{Gonzalez_et_al_2007,Skinner_et_al_2018} implemented codes that use the M1-closure approximation; in the special relativistic regime in \citep{takahashi_et_al_2013}  the matter was coupled with radiation using the Eddington approximation, later on, in \citep{takahashi_Ohsuga_2013} the authors extended the numerical method to include resistive magnetohydrodynamics (MHD) with the M1-closure relation; in the general relativistic (GR) context, \citep{Farris_et_al_2008} implemented a numerical scheme of GR-RMHD in which the Eddington approximation is employed; in \citep{Zanotti_et_al_2011} and \citep{Fragile_et_al_2012} an implementation of the GR-RMHD equations is used to study, in the optically thick regime, the Bondi-Hoyle accretion onto black holes and also to model the radiation in accretion disks; the M1-closure approach also was implemented in GR codes, by \citep{Roedig_et_al_2012} where the GR-RMHD was used to model more realistic temperatures of the supersonic Bondi Hoyle Lyttleton accretion in two dimensions; in \citep{Sadowski_et_al_2013b} and \citep{Fragile_et_al_2014} the robustness of two  GR-RMHD codes were presented. 

Concerning the difficulties of solving matter fields coupled to  radiation fields, one finds that besides the moment-closure relation, there is another important ingredient, namely, the opacities. This property has a considerable influence on the dynamics of the  system and if the opacity has a large value, the system of equations may become stiff. Then, evolution methods used to solve the radiation hydrodynamics equations are conditioned to use small time steps in order to preserve numerical stability \citep{Farris_et_al_2008,Zanotti_et_al_2011,Fragile_et_al_2012}.
Recently, in order to avoid instabilities due the stiffness, the state of the art numerical simulations use hybrid implicit-explicit (IMEX) schemes to evolve the system without the  time step size restriction \citep{Roedig_et_al_2012,takahashi_et_al_2013,takahashi_Ohsuga_2013,Sadowski_et_al_2013b,Fragile_et_al_2014}.

In this paper, we present {\it CAFE-R}, a  numerical code that takes into account the dynamical feedback between the matter and radiation pressures by solving the special relativistic radiation hydrodynamics equations under the Gray-Body approximation. The radiation field is solved using the moment equations of the Boltzmann equation associated with the transport of photons, using both Eddington and M1-closure approximations. We verify the numerical results using a set of established test problems in different regimes, covering from a non-relativistic gas-pressure-dominated scenario until a highly-relativistic radiation-pressure-dominated limit, in the optically thin and thick regimes. Also, we show the self-converge factor of  1D test problems. 

The objective of {\it CAFE-R} is the simulation of high-energy astrophysical scenarios in which the matter and radiation are strongly coupled, but magnetic fields and gravitational field effects are negligible. 
Among these scenarios, there are a number of interesting scenarios, including jets from AGNs \citep{Cielo_et_al_2014,Karen_Reynolds_2016}, microquasars \citep{Bosch-Ramon_Khangulyan_2009,Perucho_2012}, blazars \citep{Fromm_et_al_2016,Rueda-Becerril_et_al_2017}, GRBs \citep{Nagakura_et_al_2011,Cuesta_et_al_2015,Cuesta_et_al_2015b,Lopez-Camara_et_al_2016,DeColle_et_al_2017,Rivera_Guzman_2018}, and jets launched from the common envelope phase of two compact objects \citep{Moreno_et_al_2017,Lopez-Camara_et_al_2018} can be modeled with relativistic radiation hydrodynamics (RRHD). The special relativistic regime of our code, is appropriate in a region far from the central engine. On the other hand, the radiation plays an important role due to acceleration effects on jets  \citep{Takeuchi_et_al_2010}.  

The above scenarios seem to have in common that they are launched from a progenitor such as a compact object, as well as, they present collimated shapes and relativistic speeds. In these cases, some differences are expected if temporal and spatial scales that must be taken into account in the RRHD models vary significantly. For instance, the time scale for GRB jets is of the order of seconds, whereas for  AGN jets it is of  order of years \citep{Bottcher_et_al_2012}.  
A major advantage of solving the coupled system of matter and radiation is that it allows the construction of Light Curves directly from the radiation flux on the fly and is one of the reasons for the implementation of such schemes.
Another advantage of {\it CAFE-R} is that it also allows to distinguish the temperatures of the fluid and radiation separately, since local thermal equilibrium is assumed only at initial time . This is so important because, through  macroscopic quantities like temperature, we can infer information about the progenitor of the jet,  the nature of the surroundings of the progenitor  and  the dominant radiative processes.

This work is organized as follows. In Section \ref{sec:numerical_methods}, we show the system of equations of the special relativistic radiation hydrodynamics. In Section \ref{sec:Numerical_methods} we describe the numerical methods used to solve such equations. In Section \ref{sec:tests}, we show the standard 1D and 2D numerical tests and  self-convergence. In Section \ref{sec:jets}, as a 3D application, we present the simulation of a jet in the radiation pressure dominated regime with a helical perturbation. Finally, in Section \ref{sec:final_comments}, we discuss the potential and limitations of  {\it CAFE-R}.

\section{System of equations}
\label{sec:numerical_methods}

The equations of radiation hydrodynamics describe the evolution of a fluid coupled to a radiation field. The fluid evolution is determined by the conservation equations of mass, momentum, and energy of the fluid, which are coupled to the radiative transfer equations through source terms, which characterize the momentum and energy exchanges between the fluid and the radiation. We now describe the equations of the fluid-radiation system, for which in what follows, we assume the speed of light is one, Latin indices range over 1-3, and Greek ones do over 0-3, where the 0 corresponds to the temporal component and the 1-3 represent the spatial components of the tensors involved. The equations that govern the evolution of the system are:

\begin{eqnarray}
	\label{eq:hy}
\nabla_\alpha(\rho u^\alpha) &=&0,\\
\nabla_\alpha T^{\alpha\beta}_\text{m} &=&G^\beta_\text{r},\\
\nabla_\alpha T^{\alpha\beta}_\text{r} &=&-G^\beta_\text{r},
\label{eq:rad}
\end{eqnarray}

\noindent with $\rho$ the rest-mass density, $u^\alpha$ the four-velocity of fluid elements and $T^{\alpha\beta}_\text{m}$ is the stress energy tensor of the fluid

\begin{equation}
T^{\alpha\beta}_\text{m} = \rho hu^\alpha u^\beta + Pg^{\alpha\beta},
\end{equation}

\noindent where $g^{\alpha\beta}=(-1,1,1,1)$ is the metric of Minkowski space-time, $h=1+\epsilon+P/\rho$ is the specific enthalpy, $\epsilon$ the specific internal energy, and $P$ the thermal pressure. The thermal pressure is related to $\rho$ and $\epsilon$ through a gamma-law equation of state (EoS) $P=\rho\epsilon(\Gamma-1)$, where $\Gamma$ is the adiabatic index of the gas.  Finally, $T^{\alpha\beta}_\text{r}$ is the stress-energy tensor that describes the radiation field in (\ref{eq:rad}). We treat the radiation in the laboratory frame, rather than in the comoving frame. This approach was introduced in \citep{takahashi_et_al_2013,takahashi_Ohsuga_2013} and is slightly different from the approach in our previous applications  \citep{Rivera_Guzman_2016,Rivera_Guzman_2018}. In the laboratory frame this tensor reads
 
\begin{equation}
T^{\alpha\beta}_\text{r} = \begin{pmatrix} 
E_\text{r} & F^j_\text{r} \\
F^i_\text{r} & P^{ij}_\text{r} 
\end{pmatrix},
\end{equation}

\noindent where $E_\text{r}$ is the density of radiated energy, $F^i_\text{r}$ is the radiation flux, and $P^{ij}_\text{r}$ is the radiation pressure tensor,  quantities that correspond to the zeroth, first, and second moments of the Boltzmann equation, respectively. All of them are measured in the laboratory frame. This  frame is useful because the equations for radiation  are  hyperbolic, a  convenient property for numerical integration. However, a drawback is that, using the radiative quantities expressed in the comoving frame adds non-conservative terms to the equations, and  transformations between the comoving and laboratory frames are required in order to calculate  observables. Then,  interactions between radiation and matter become complex because of Doppler and aberration effects that have to be incorporated in the source terms $G^{\alpha}_\text{r}$. 

An important assumption in {\it CAFE-R} is that the radiation field obeys the grey-body (GB) approximation, then the source terms read explicitly as \citep{takahashi_et_al_2013}

\begin{eqnarray}
	G^0_\text{r} &=& -\chi^t(4\pi BW - WE_\text{r} + u_iF^i_\text{r})\label{eq:G0}  \\
	\nonumber
	      & & - \chi^s[Wu^2E_\text{r} + Wu_iu_jP^{ij}_\text{r}-(W^2+u^2)u_iF^i_\text{r}],
\nonumber\\
    G^i_\text{r} &=& -\chi^t 4\pi Bu_i + (\chi^t + \chi^s)(WF^i_\text{r} u_jP^{ij}_\text{r}) \label{eq:Gi}\\
	      & & -\chi^su_i(W^2E_\text{r} - 2Wu_jF^j_\text{r} +u_ju_kP^{jk}_\text{r}),\nonumber
\end{eqnarray}

\noindent where $u=\sqrt{u_iu^i}$, $W$ is the Lorentz factor, and $\chi^t =\kappa^t\rho$, $\chi^s=\kappa^s\rho$ are the opacity coefficients, where the superscripts $t$ and $s$ denote the thermal and scattering opacities respectively. The GB assumption means that the $\kappa's$ are  independent of the frequency, then, the Planck function is $B=\frac{1}{4\pi}a_\text{r}T^4_\text{fluid}$, with $T_\text{fluid}$ the temperature of the fluid, and $a_\text{r}$ the radiation constant. It is important to point out that the source terms (\ref{eq:G0}) and (\ref{eq:Gi}) become stiff in regimes where the optical thickness is high.

In order to close the  system of equations, an extra condition that relates the second moment of radiation with one of the lower order moments is needed. The simplest approach is the Eddington approximation, which assumes a nearly isotropic radiation field and in the comoving frame it shows a pressure tensor with the form \citep{Mihalas_Mihalas_1984}

\begin{equation}
P^{ij}_\text{r,co} = \frac{1}{3}E_\text{r,co}\delta^{ij},
\label{eq:Edd}
\end{equation} 

\noindent where subindex ``co'' indicates the quantity is defined in the comoving frame. This assumption is valid only in the optically thick regime within the diffusion limit. In order to obtain the radiation pressure $P^{ij}_\text{r}$ in the laboratory frame, a Lorentz transformation on the radiation energy momentum tensor is needed, specifically the radiation stress tensor in the laboratory  and in the comoving frames are related through the following equation \citep{Myeong-Gu_2006}

\begin{eqnarray}
\label{eq:LorentzPr}
P^{ij}_\text{r}  &=&  W^2v^iv^jE_\text{r,co} + W\left(v^i\delta^j_k + v^j\delta^i_k -2\frac{W-1}{v^2}v^iv^jv_k\right)F^{k}_\text{r,co} \\
\nonumber
                   & & +\left(\delta^i_k +\frac{W-1}{v^2}v^iv_k\right)\left(\delta^j_k +\frac{W-1}{v^2}v^jv_l\right)P^{kl}_\text{r,co}.
\end{eqnarray}

\noindent Substituting  Eq. (\ref{eq:Edd}) into Eq. (\ref{eq:LorentzPr}), together with the respective Lorentz transformation on the zeroth and first moments of radiation, one obtains the explicit form of the radiation pressure in the laboratory frame   \citep{takahashi_et_al_2013} 

\begin{eqnarray}
\label{eq:Pr}
P^{ij}_\text{r} &+& \left[ -\frac{\delta^{ij}}{3} + \frac{W^2v^iv^j}{(i +W)^2}\right]W^2v_kv_mP^{km}_\text{r} \\
\nonumber
                &+&  \frac{W}{1 + W}(v^iv_kP^{ik}_\text{r} + v^jv_kP^{jk}_\text{r}) = R^{ij},
\end{eqnarray}

\noindent where

\begin{eqnarray}
\label{eq:R}
R^{ij} &=& \frac{\delta^{ij}}{3}(E_\text{r} - 2v_kF^k_\text{r})W^2 - W^2v^iv^jE_\text{r}  \\
\nonumber
       &+&  W^2(v_iF^j_\text{r} + v_jF^i_\text{r}) + \frac{W^3}{1 + W}v^iv^jv_kF^{k}_\text{r}.
\end{eqnarray}

In order to find all the individual components of the radiation pressure tensor, we need to solve  Eq. (\ref{eq:Pr}) numerically. Following the strategy in \citep{takahashi_et_al_2013}, we write  Eqs. (\ref{eq:Pr}-\ref{eq:R}) as

\begin{equation}
{\cal A}(v){\cal P}={\cal R},	
\end{equation}  

\noindent where ${\cal A}(v)$ is a $6\times 6$ matrix that depends only on the fluid velocity, ${\cal P}^T=(P^{11}_\text{r},P^{22}_\text{r},P^{33}_\text{r},P^{12}_\text{r},P^{13}_\text{r},P^{12}_\text{r})$ and ${\cal R}^T=(R^{11},R^{22},R^{33},R^{12},R^{13},R^{12})$. 
We compute the components $P^{ij}_\text{r}$ by inverting the matrix ${\cal A}(v)$ using the LU-decomposition.

The Eddington approach is appropriate when the radiation field is well coupled with the fluid. However, when the radiation field and the fluid are weakly coupled, a more general assumption is required to have a good approximation. A scheme that allows a description of the radiation field in both, optically thick and thin regimes is the M1-closure \citep{Levermore_1984,Dubroca_Feugeas_1999,Gonzalez_et_al_2007,takahashi_Ohsuga_2013}. The M1-closure provides a better approximation than Eddington to the radiation field because it describes the diffusion limit, as well as the free-streaming limit where the radiative energy is transported at the speed of light. This closure relation is given by

\begin{equation}
P^{ij}_\text{r} = \left(\frac{1-\zeta}{2}\delta^{ij} + \frac{3\zeta-1}{2}n^in^j\right) E_\text{r},
\label{eq:pressuretensor}
\end{equation}  

\noindent where $n^i = F^i_\text{r}/|{\bf F_\text{r}}|$, $\zeta=\frac{3+4|{\bf f}|^2}{5+2\sqrt{4-3|{\bf f}|^2}}$ is the Eddington factor \citep{Levermore_1984}, the expression in parentheses is the Eddington tensor, and $f^i=F^i_\text{r}/cE_\text{r}$ is the reduced radiative flux. Notice that the Eddington factor of this model is a function of $E_\text{r}$ and $F^i_\text{r}$, which can be evaluated in the laboratory frame. Thus, we can directly obtain $P^{ij}_\text{r}$ from $E_\text{r}$ and $F^i_\text{r}$ without any Lorentz transformation. The M1-closure relation contains the optically thick and optically thin regimes. In the optically thick regime $F_\text{r}^i\approx 0$, $f^i=0$, and $\zeta=1/3$, that corresponds to Eddington's approximation. On the other hand, in the optically thin regime $F_\text{r}^i= cE_\text{r}$, $f^i=1$, and $\zeta=1$, that is associated to the free-streaming limit. 

The gas temperature is estimated from the ideal-gas EoS via the expression $T_\text{fluid} = \frac{\mu m_p}{k_B}\frac{P}{\rho}$, being $k_B$ the Boltzmann constant, $\mu$ the mean molecular weight and $m_p$ the proton mass, that we assume in this paper to be $\mu=1$ in all cases. This is a good approximation only when the fluid pressure is much greater than the radiation pressure and/or when the radiation field is weakly coupled with  matter; otherwise, the temperature must be calculated taking into account the contributions of  baryons, radiation pressure, and optical depth. An approximate expression that captures the effects of these two regimes of radiative transfer for the total pressure, similar to that in \citep{Cuesta_et_al_2015} is

\begin{equation}
P_\text{t} = \frac{k_B}{\mu m_p}\rho T_\text{fluid} + (1 - e^{-\tau})\zeta(T_\text{rad}) a_\text{r}T_\text{rad}^4,
\label{eq:tem}
\end{equation}

\noindent where $\tau$ is the  optical depth computed at each numerical cell of the domain, along  straight lines parallel to each direction $ \hat{x},~\hat{y},~\text{or}~\hat{z}$.  The difference between (\ref{eq:tem}) and the relation in \citep{Cuesta_et_al_2015} is that $\zeta$ depends on the temperature of the radiation, which is consistent with the closure M1 in (\ref{eq:pressuretensor}). Since the optical depth of a relativistic moving medium strongly depends on its velocity and on the viewing angle \citep{Abramowicz_et_al_1991}, one needs to compute this variable taking into account the Doppler effect  

\begin{equation}
\tau=\int(\chi^t+\chi^s)W(1 - |{\bf v}|\cos \theta)ds,	
\end{equation}

\noindent where $\cos \theta = s/(x^2 + y^2 + z^2)^{1/2}$, and $T_\text{rad}=(E_\text{r}/a_\text{r})^{1/4}$ is the temperature of radiation. Here, $\tau$ depends on the temperature only if any of the opacity coefficients does. When the fluid and radiation are in local thermal equilibrium (LTE), that is  $T_\text{fluid}=T_\text{rad}$, the temperature approximately obeys a fourth order equation similar to Eq. (\ref{eq:tem}) above \citep{Cuesta_et_al_2015}. In the general case, in order to compute  the fluid temperature from Eq. (\ref{eq:tem}), we use a Newton-Raphson method under the assumption of LTE for the initial guess of temperature. Notice that in purely hydrodynamical models, the temperature is an auxiliary quantity,  whereas in radiation hydrodynamics models, it is essential because it may substantially change the opacities, and therefore the behavior of the whole system. From Eq. (\ref{eq:tem}), we can see that the value of the optical depth influences importantly the fluid temperature.

It is important to mention that there is another more general EoS (Helmholtz EoS), which not only includes contributions  from a radiation field and baryons, but also for completely ionized nuclei,  degenerate relativistic electrons and positrons \cite{Timmes_Swesty_1999}. 

\section{Numerical methods}
\label{sec:Numerical_methods}

So far, we  have presented the RRH time-dependent evolution equations that model the radiative transfer on a moving fluid. Now, we describe the numerical methods implemented in  {\it CAFE-R}  used to solve the RRH equations. First, we write the RRH equations in the flux balance form $\partial_0\textbf{U}+\partial_i\textbf{F}^i=\textbf{S}$ using Cartesian coordinates in 3D, where  $\textbf{U}$ is the vector of conserved variables, which  are functions of the primitive variables  $\textbf{p}^T=(\rho,v^i,P,E_\text{r},F^i_\text{r})$, $\textbf{F}^i$ are the fluxes and  $\textbf{S}$ the sources, which are given by

\begin{equation}
 \textbf{U} = \begin{bmatrix}
      D  \\[0.3em]
      S^i \\[0.3em]
      \tau \\[0.3em]
      \tau_\text{r}\\[0.3em]
      S^i_\text{r}
     \end{bmatrix}
       = \begin{bmatrix}
       \rho W \\[0.3em]
       \rho hW^2v^i\\[0.3em]
       \rho hW^2-P-\rho W\\[0.3em]
       E_\text{r}\\[0.3em]
       F^i_\text{r}
     \end{bmatrix},\label{eq:consvars}
\end{equation}
\begin{equation}
 \textbf{F}^i = \begin{bmatrix}
      v^iD  \\[0.3em]
       S^jv^i + P\delta^{i,j}\\[0.3em]
      \tau v^i + Pv^i \\[0.3em]
      F^i_\text{r}\\[0.3em]
      P^{ji}_\text{r}
     \end{bmatrix},
\ \ \textbf{S} =\begin{bmatrix}
       0 \\[0.3em]
       G^i_\text{r}\\[0.3em]
       G^0_\text{r}\\[0.3em]
       -G^0_\text{r}\\[0.3em]
       -G^i_\text{r}
     \end{bmatrix}. 
\end{equation}

\noindent Notice that the conservative and primitive variables of radiation are the same $(E_\text{r},F^i_\text{r})$. Thus, the conversion between radiation conserved and primitive quantities is straightforward.
The coupling of radiation happens through the sources (\ref{eq:G0}) and (\ref{eq:Gi}), which require the calculation of $T_{\rm fluid}$, which is obtained and used as follows. We recover the primitive hydrodynamical variables by solving the typical transcendental equation for the fluid pressure $P$ that depends on $D,S^i,\tau$. With this $P$ and $\rho=D/W$ it is possible to use (\ref{eq:tem}) to obtain $T_{\rm fluid}$. Then this fluid temperature is inserted in the sources (\ref{eq:G0}) and (\ref{eq:Gi}) that couple fluid and radiation. This coupling is practiced at every step and intermediate step during the time integration. 
Since this coupling is essential to the appropriate implementation, at this point we want to enhance the use of Eq. (\ref{eq:tem}) with the following observation. It is possible to compute the fluid temperature by equating the pressure obtained from the gamma-law EoS $P=\rho\epsilon(\Gamma-1)$ with that of the ideal gas $T_{\rm fluid}=\frac{\mu m_p}{k_B}\frac{P}{\rho}$, that is $T_{\rm fluid,\Gamma}=\frac{\mu m_p}{k_B}\epsilon(\Gamma-1)$ which is the part of Eq. (\ref{eq:tem}) related exclusively to the fluid, that is, this formula {\it does not} involve the radiation effects. In fact the expression for $T_{\rm fluid,\Gamma}$ and  (\ref{eq:tem}) coincide when $P^{ij}_{\rm r}\simeq 0$, namely, in scenarios where the fluid pressure dominates over radiation pressure. In the appendix we present examples of the calculation of fluid temperature using  (\ref{eq:tem}) and $T_{\rm fluid,\Gamma}$ in hydrodynamical and radiation pressure dominated cases.

For the solution of the system of equations we use the method of lines with uniform space and time resolutions related by $\Delta t = CFL \cdot \min(\Delta x,\Delta y,\Delta z)$, where CFL is the Courant-Friedrichs-Lewy factor. 
For the  integration in time we use a second order accurate IMEX Runge-Kutta integrator, in which the hydrodynamical variables are solved explicitly whereas the radiation variables are solved implicitly, following the strategy in  \cite{Roedig_et_al_2012}, where the right hand side of the equations is split into two parts

\begin{equation}
\partial_0 \textbf{U}= \textbf{H}(\textbf{U}) + \textbf{K}(\textbf{U}),
\label{eq:split}
\end{equation}

\noindent where \textbf{H} is an operator that contains the spatial derivatives of the conserved hydrodynamical variables $\{D,S^i,\tau\}$ and its respective source terms $\{0,G^i,G^0\}$. On the other hand, \textbf{K} will be defined by the spatial derivatives of the radiation conservative variables $\{\tau_\text{r},S^i_\text{r}\}$  and the source terms $\{-G^0,-G^i\}$. Keeping this in mind, we implemented a particular solution of Eq.(\ref{eq:split}) based on those presented in \citep{Higueras_2006}, in which the construction of $U^{n+1}$ is given by

\begin{eqnarray} 
\textbf{U}^*     &=& \textbf{U}^n + \frac{dt}{2}[\textbf{H}(t^n + \frac{1}{2}dt,\textbf{U}^n) + 
\nonumber\\
        & & \textbf{K}(t^n + \frac{1}{2}dt,\textbf{U}^*)],\label{eq:implicitpart}\\
\textbf{U}^{n+1} &=& \textbf{U}^n + dt[\textbf{H}(t^n + \frac{1}{2}dt,\textbf{U}^*) + 
\nonumber\\
        & & \textbf{K}(t^n + \frac{1}{2}dt,\textbf{U}^*)].\label{eq:explicitpart}
\end{eqnarray}

\noindent Notice that the variable $\textbf{U}^*$ appears in both sides of (\ref{eq:implicitpart}), as usual in implicit methods, which defines an algebraic equation for this variable that we solve using a Newton-Raphson method.

The spatial part is constructed with a finite volume discretization and the numerical fluxes at the space cell interfaces are computed with a high-resolution shock-capturing method that uses the HLLE numerical flux formula. This formula along the $j-$direction, where $j$ labels the $x,y,z$ axes, is given by

\begin{equation}
\textbf{F}^{i-\text{HLLE}}_{j+1/2}= \frac{\lambda^{+}\textbf{F}^i(\textbf{U}^\text{L}_{j+1/2}) - \lambda^{-}\textbf{F}^i(\textbf{U}^\text{R}_{j+1/2}) + \lambda^{+}\lambda^{-}(\textbf{U}^\text{R}_{j+1/2} - \textbf{U}^\text{L}_{j+1/2})}{\lambda^{+} - \lambda^{-}},	
\end{equation}

\noindent where $\textbf{U}^\text{R}_{j+1/2}$ and $\textbf{U}^\text{L}_{j+1/2}$ are the values of the conservative variables reconstructed at the right and left from the intercell boundary, respectively. The minmod- and mc-slope limiters are used for the intercell reconstruction of the conserved variables. The wave velocities are computed from the eigenvalues of the Jacobian matrix $J^i = \partial \textbf{F}^i / \partial \textbf{U}$. Finally, $\lambda^+$ and $\lambda^-$ are the fastest and slowest among the characteristic wave velocities of the system respectively. 

Since the radiative and hydrodynamical variables of the system of equations (\ref{eq:hy}-\ref{eq:rad}) are coupled only through the source terms $G^{\alpha}_\text{r}$, we calculate the radiation and hydrodynamics wave speeds separately. This means that the Jacobian matrix $J^i$ can be separated into two sub-matrices, one for the hydrodynamical variables $J^i_\text{hyd}$ and another one for the radiation field $J^i_\text{rad}$. Thus, we can compute the eigenvalues of hydrodynamics and radiation equations independently  \citep{Sadowski_et_al_2013b}. Specifically, the five  eigenvalues of $J^i_\text{hyd}$ are those of the purely relativistic hydrodynamics, for instance  along the $\hat{x}-$direction these are

\begin{eqnarray}
	\nonumber
	\lambda_1 &=& v^x, \\
	\nonumber
	\lambda_{2,3} &=& \left( v^x(1-c_s^2)\pm c_s\sqrt{[1-v^2][1-v^2c_s^2 - v^xv^x(1-c_s^2)]} \right) \\
	 \nonumber
	              & &\left(\frac{1}{1-v^2c_s^2}\right),
\end{eqnarray}
	
\noindent where $\lambda_1$ is a triply degenerated eigenvalue, $v^2=v^iv^j$, and $c_s =\sqrt{P\Gamma/(h\rho)}$ is the sound speed. The eigenvalues in the $\hat{y}-,~\hat{z}-$directions are similar to the above eigenvalues and they can be obtained from an adequate index permutation \citep{Font_et_al_1994}.

On the other hand, the Jacobian matrix for the radiation part is

\begin{equation}
J^i_\text{rad} = \begin{pmatrix} 
\partial F^i_\text{r}/\partial E_\text{r}    & \partial F^i_\text{r}/\partial F^j_\text{r} \\
\partial F^{ij}_\text{r}/\partial E_\text{r} & \partial P^{ij}_\text{r}/\partial F^j_\text{r}
\end{pmatrix},
\end{equation}
 
\noindent whose eigenvalues depend on the closure relation. In the Eddington approximation the characteristic speeds are $\pm 1\sqrt{3}$, whereas in the M1 model, the characteristic velocities are functions of $E_\text{r}$ and $F^i_\text{r}$, that specifically depend on the norm of the reduced flux ($|{\bf f}|$) and on the angle $\theta$ that $n^i=f^i/|{\bf f}|$ makes with the interface normal to the $i-$direction, where $i$ labels each of the Cartesian coordinates $x,~y,~ \text{or}~ z$. 
In order to obtain explicit formulas for the eigenvalues, following \citep{Skinner_Ostriker_2013}, we rotate the local system of coordinates around $\hat{x}$, changing from $(x,y,z)$ to $(x,y^\prime,z^\prime)$, so that $\hat{z}^\prime\cdot n^i=0$ in the prime coordinate system. In this new coordinate system, the three -out of four- linearly independent eigenvalues are

\begin{eqnarray}
	\lambda_1 &=& \cos\theta \left[ \frac{2-\sqrt{4-3|{\bf f}|^2}}{|{\bf f}|} \right], \nonumber\\
	\lambda_{2,3} &=& \Biggl[|{\bf f}|\cos\theta\pm \Bigl[ \frac{2}{3}\left(4-3|{\bf f}|^2 - \sqrt{4-3|{\bf f}|^2}\right) \nonumber\\
	              &+& 2\cos^2\theta\left(2-|{\bf f}|^2 - \sqrt{4-3|{\bf f}|^2}\right)\Bigl]^{1/2} \Biggl] \nonumber\\
	           &\div& \sqrt{4-3|{\bf f}|^2},  \nonumber
\end{eqnarray}

\noindent where $\cos\theta=\hat{x}\cdot n^i=f^x/\sqrt{f^z+f^y+f^z}$. For the optically thin limit ($f\rightarrow 1$), the characteristic speeds are $\lambda_{1,2,3}\rightarrow \hat{x}\cdot n^i$. Particularly when $\hat{x}$ and $n^i$ are parallel, the regime in which the fastest characteristic velocity is the speed of light is recovered. When $\hat{x}$ and $n^i$ are perpendicular, there is no longer transport along the $x-$direction. Furthermore, for the optically thick limit ($f\rightarrow 0$), the eigenvalues are $\lambda_1\rightarrow 0$ and $\lambda_{2,3}\rightarrow \pm 1/\sqrt{3}$. These should correspond to the fastest characteristic velocity given by the diffusion theory.  However, when $\tau\gg1$, the signal speeds can be overestimated, that means that $\lambda_{2,3}$ have values numerically larger than $1/\sqrt{3}$, causing  additional numerical diffusion. To avoid this numerical diffusion, we follow the suggestion made by \cite{Sadowski_et_al_2013b}, which consists in modifying the characteristic velocities along each Cartesian direction as

\begin{eqnarray}
	\lambda^+_\text{co} &=& \text{min}\left(\lambda^+_\text{co}, \frac{1}{3\tau^i}\right), \\
    \lambda^-_\text{co} &=& \text{max}\left(\lambda^-_\text{co}, -\frac{1}{3\tau^i}\right),
\end{eqnarray}

\noindent where $\tau^i$ is the optical depth in a cell and $\lambda^\pm_\text{co}$ are the characteristic velocities in the comoving frame. This modification allows the reduction of  numerical diffusion in the optically thick regime.

Finally, it is important to point out that unphysical solutions appear when $|F^i_\text{r}| > E_\text{r}$. To guarantee that the constraint $|F^i_\text{r}| \leq E_\text{r}$ is satisfied in our numerical scheme, we modify the radiation flux as

\begin{eqnarray}
	F^i_\text{r} &=& F^i_\text{r} \text{min}\left(1, \frac{E_\text{r}}{|F^i_\text{r}|}\right),
\end{eqnarray}

\noindent which reduces the radiation flux without changing its direction.

\section{Tests}
\label{sec:tests}

	\begin{table*}[]	
	\begin{center}
	\begin{tabular}{c  c  c  c  c  c  c  c  c  c  c  c}
	\hline
	\hline
	Test &$\Gamma$ & $a_\text{rad}$ & $\kappa^\text{t}$ & $\rho_\text{L}$ & $P_\text{L}$ & $u^x_\text{L}$& $E_\text{r,co,L}$ & $\rho_\text{R}$ & $P_\text{R}$ & $u^x_\text{R}$ & $E_\text{r,co,R}$ \\
	\hline
	 $1$ & 5/3 &$1.234\times10^{10}$&$0.4$&$1.0$&$3.0\times10^{-5}$&$0.015$&$1.0\times10^{-8}$&$2.4$ &$1.61\times10^{-4}$&$6.25\times10^{-3}$&$2.51\times10^{-7}$\\
	\hline
	 $2$ & 5/3&$7.812\times10^{4}$&$0.2$   &$1.0$&  $4.0\times10^{-3}$&$0.25$ &  $2.0\times10^{-5}$ &$3.11$&$4.51\times10^{-2}$&$0.0804$&$3.4\times10^{-3}$\\
	\hline
	 $3a$ & 2 &$1.543\times10^{-7}$&$0.3$&$1.0$&  $60.0$          &$10.0$ &  $2.0$           &$8.0$ &$2.34\times10^{3}$&$1.25$&$1.14\times10^{3}$\\
     \hline
     $3b$ & 2   &$1.543\times10^{-7}$&$25$&$1.0$&  $60.0$          &$10.0$ &  $2.0$           &$8.0$ &$2.34\times10^{3}$&$1.25$&$1.14\times10^{3}$\\
	 \hline
	 $4a$ & 5/3 &$1.388\times10^{8}$&$0.08$&$1.0$ &$6.0\times10^{-3}$&$0.69$ &  $0.18$          &$3.65$&$3.59\times10^{-2}$&$0.189$ & $1.3$\\
     \hline
     $4b$ & 5/3 &$1.388\times10^{8}$&$0.7$&$1.0$ &$6.0\times10^{-3}$&$0.69$ &  $0.18$          &$3.65$&$3.59\times10^{-2}$&$0.189$ & $1.3$\\
     \hline
     $5$ & 2   &$1.543\times10^{-7}$&$1000$&$1.0$&  $60.0$          &$1.25$ &  $2.0$           &$1.0$ &$60.0$&$1.10$&$2.0$\\
	\hline
	\hline
	\end{tabular}
	\caption{\label{table:testGB} Parameters of the  shock tube tests. $L$ and $R$ serve to label state variables to the left $x<0$ and to the right $x>0$ from the initial discontinuity. In all cases we set $\chi^s = \kappa^s =0$ and the radiation flux $F^i_\text{r,co} = 0$ initially. The velocities $u^{x}_{R}$ and $u^{x}_{L}$ are the $x-$component of the 3 velocity multiplied by the Lorentz factor.}
	\end{center}
	\end{table*}

	In order to show that our implementation works correctly, we produce the tests in \citep{takahashi_Ohsuga_2013} using both M1  and Eddington approximation closure relations. We first show 1D tests, which are basically Riemann problems corresponding to shock-tubes with different initial states designed to illustrate different scenarios, aligned along the $x-$axis with the discontinuity at $x=0$. The domain for tests 1,2 and 5 below, is $x\in[-20,20]$, whereas for tests 3 and 4 the domain is $x\in[-80,80]$, even though in the figures we present the zoomed results in a smaller domain. We use a production resolution $\Delta x=0.05$ for all the 1D tests and cover the domain along the additional $y$ and $z-$directions using 5 cells. For the reconstruction of variables we use the MC-slope limiter and time integration uses a CFL factor equal to $0.25$. Finally, the boundary conditions are outflow.   The essential initial conditions for all the problems appear in Table \ref{table:testGB}, and the initial values for the radiation flux and the scattering opacity are set to $F^i_\text{r,co} = 0$ and  $\chi^s = 0$ in all cases. Subscripts $L$ and $R$ denote the left $(x < 0)$ and right $(x > 0)$ states of the Riemann problem, respectively.\\
 
\subsection{Non-relativistic Strong Shock}

\begin{figure}
\centering
\includegraphics[width=9cm]{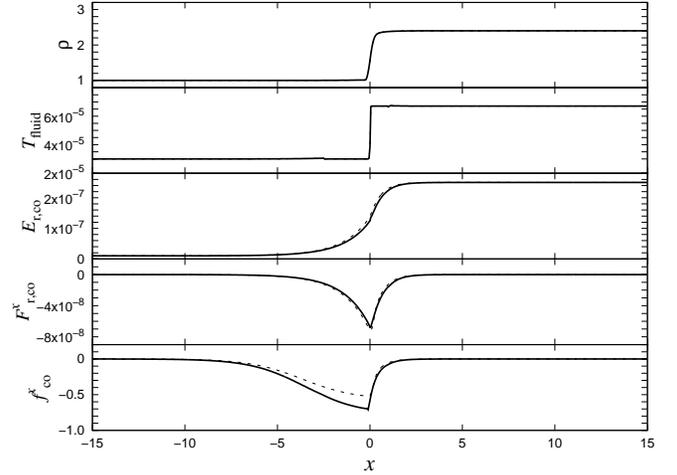}
\caption{\label{fig:test1} Snapshot of the results for Test 1 at time $t=3500$ as seen by a comoving observer. Shown are the rest-mass density and temperature of the fluid, radiation energy density,  radiation flux and the reduced radiative flux. Results for the M1 and Eddington closures correspond to solid and dashed curves.}
\end{figure}

This is Test 1 and in Figure \ref{fig:test1} we show the mass density, fluid temperature, radiation energy density,  radiation flux, and the reduced flux $f^{x}_\text{co}=F^{x}_\text{r,co}/E_\text{r,co}$ at $t = 3500$ measured in the comoving frame.

The initial conditions are set in such a way that the energy density of the fluid dominates over the radiation energy density. In consequence hydrodynamical effects are more important than those due to radiation pressure and consequently the results  are similar to those of a hydrodynamical shock-tube test. Another implication is that the results using the two closure models, Eddington and M1, show pretty similar results. For instance, the difference in temperature, radiation energy density $E_\text{r,co}$ and radiation flux $F^{x}_\text{r,co}$ profiles is small. The initial conditions imply the radiation flux is being transported from right to left, and has an exponential decrease with distance $\propto e^{-\chi^t|x|}$ and $ \propto e^{-\sqrt{3}\chi^t|x|}$ for M1 and  Eddington closures respectively as discussed in \cite{takahashi_Ohsuga_2013}. The factor $\sqrt{3}$ in the exponential is the factor in the propagation speed between the two closure methods.

Finally, in the bottom panel we compare the reduced flux for the two models, showing that for the M1 closure this quantity is slightly smoother than for the Eddington one.

\subsection{Mildly-relativistic strong shock}
	
\begin{figure}
\centering
\includegraphics[width=9cm]{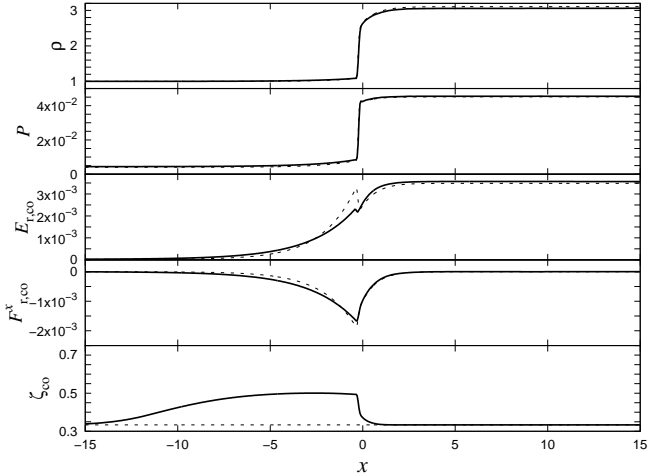}
\caption{\label{fig:test2}
Snapshot of the results for Test 2 at time $t=3500$ as seen by a comoving observer. Shown are the rest-mass density and pressure of the fluid, radiation energy density,  radiation flux and the $xx$ component of the Eddington tensor. Results for the M1 and Eddington closures we use  solid and dashed curves.}
\end{figure}
	
	This is Test 2 and corresponds to a   gas-pressure dominated strong shock. In Figure \ref{fig:test2}, we show the rest-mass density, fluid pressure, radiative energy density, radiation flux, and the $xx$ component of the Eddington tensor in the comoving frame at $t=3500$. In this test, not only the hydrodynamical variables are discontinuous, but also the radiation energy density and the radiative flux.  Our results are in concordance with results in \cite{Tolstov_et_al_2015}, who  estimated that the upper limit of the amplitude of this discontinuity is $\sim 3.5\times 10^{-4}$.  From Figure \ref{fig:test2} we can see that the result obtained with the M1-closure presents a smaller amplitude in the discontinuity than the one obtained with the Eddington approximation. This means that the M1 model, with a smaller discontinuity gives a better result for this test.
	
\subsection{Relativistic Shock}

\begin{figure}
\centering
\includegraphics[width=9cm]{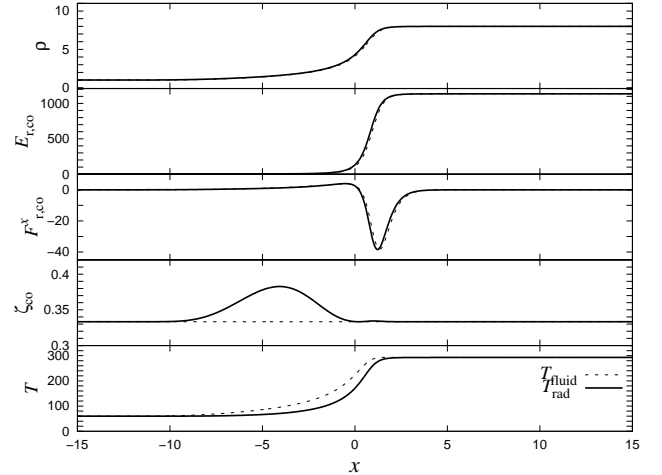}
\caption{\label{fig:test3a} 
Snapshot of the results for Test $3a$ at time $t=3500$ as seen by a comoving observer. Shown are  the rest-mass density of the fluid, radiation energy density,  radiation flux and  the $xx$ component of the Eddington tensor. For these four quantities the results for the M1 and Eddington closures use solid and dashed curves. At the bottom we show the fluid and radiation temperatures.}
\end{figure}

\begin{figure}
\centering
\includegraphics[width=9cm]{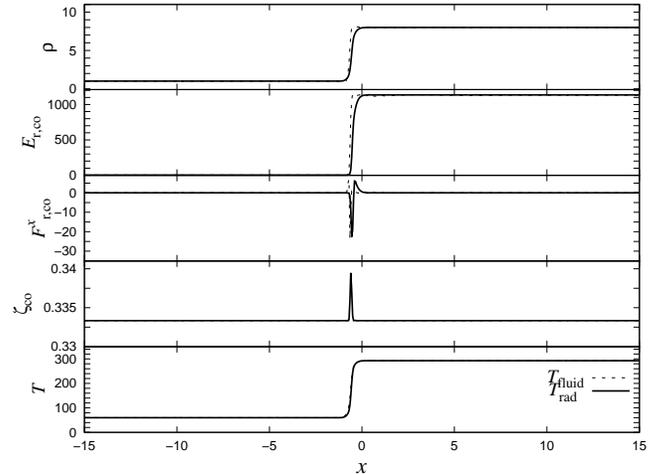}
\caption{\label{fig:test3b} Snapshot of the results for Test $3b$ at time $t=3500$ as seen by a comoving observer. Shown are  the rest-mass density of the fluid, radiation energy density,  radiation flux and  the $xx$ component of the Eddington tensor. For these four quantities the results for the M1 and Eddington closures correspond to solid and dashed curves. At the bottom we show the fluid and radiation temperatures.}
\end{figure}

This is Test 3$a$,  which has a Lorentz factor near  $10$. 
A snapshot of the fluid density, radiation energy density, radiation flux and the $xx$ component of the Eddington tensor  at $t=3500$ appears in Figure \ref{fig:test3a}. 
In agreement with results in  \citep{takahashi_Ohsuga_2013}, the shock location does not change for the Eddington case, however when using M1 it approaches  a drifted  stationary state with an approximate velocity of $\sim1.02\times 10^{-4}$. 
The drifting of the shock can be due to leakage of radiation flux through the boundaries that difficults keeping the initial two states constant, and both the leakage and drifting velocity reduce by pulling the boundaries further out, which is the reason to use the domain $x\in[-80,80]$ in this and the following three tests. Due to the drifting, the comparison of the two results uses a relocation of the  shock for the M1 case. Additionally,  in this figure we show the temperature of the fluid and radiation for the M1 closure.
The gas density, radiation energy and flux are similar using the two models. This is due to the large optical depth a scenario consistent with the Eddington closure. Nevertheless, the Eddington factor $\zeta_\text{co}$ is considerably different, and departs from the value 1/3 when using the M1 model. This indicates the radiation field is anisotropic which is in agreement with  \citep{takahashi_Ohsuga_2013} $\zeta_\text{co}\sim 0.38$. Also, we obtain that the fluid temperature is higher than that of radiation, which indicates the energy transfers from the fluid to the radiation field.

Test 3$b$ is the optically thick version of this case with a high value of $\chi^t$ and the results are shown in Figure \ref{fig:test3b}. Similar to Test $3a$, we show the profiles of the rest-mass density,  radiation energy density,  radiative flux and the $xx$ component of the Eddington tensor, and, in the bottom panel, the fluid and radiation temperatures, only for the M1 model, at $t = 3500$.
	
In Test 3$b$ the front of the shock, for the M1 model drifts with a velocity of $v_\text{sh} \sim 8 \times 10^{-5}$   in code units, which is slower than the shock front of Test 3$a$. This is because the optical depth  of Test 3$b$ is higher than that of Test 3$a$, and then it reduces the dynamics of the system. Due to the high optical  thickness in Test 3$b$, the M1 model deviates from the Eddington approach by $\sim 2\%$, see $\zeta_\text{co}$ in Figure \ref{fig:test3b}. Notice that the fluid and radiation temperature are practically the same, this means that the system is in near LTE, which occurs in regions where the optical depth is large.
	
	These two tests, Tests 3$a$ and 3$b$, verify that the code is able to resolve a highly relativistic wave in two different optical depth regimes.

\subsection{Radiation Pressure-dominated Shock}

\begin{figure}
\centering
\includegraphics[width=9cm]{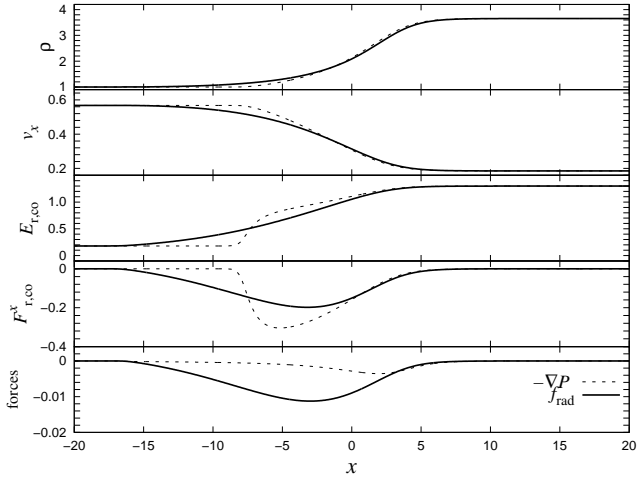}
\caption{\label{fig:test4a} 
Snapshot of the results for Test $4a$ at time $t=3500$ as seen by a comoving observer. Shown are  the rest-mass density and velocity of the fluid, radiation energy density and the radiation flux. For these four quantities the results for the M1 and Eddington closures correspond to solid and dashed curves. At the bottom we show the forces acting on the system, namely the radiative force and the fluid's gradient pressure.}
\end{figure}

\begin{figure}
\centering
\includegraphics[width=9cm]{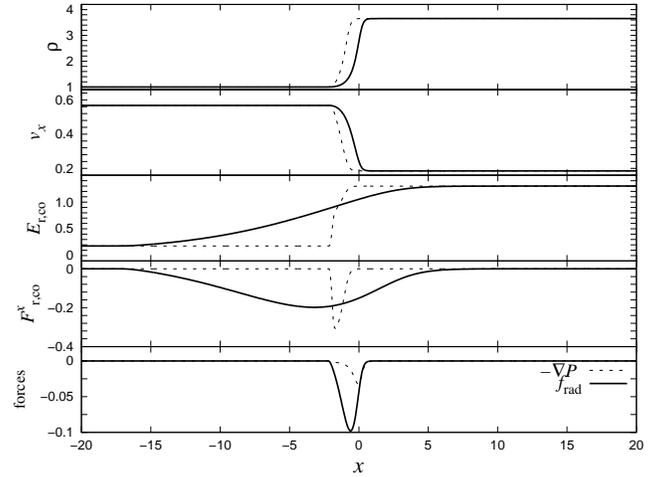}
\caption{\label{fig:test4b} Snapshot of the results for Test $4a$ at time $t=3500$ as seen by a comoving observer. Shown are  the rest-mass density and velocity of the fluid, radiation energy density and  radiation flux. For these four quantities the results for the M1 and Eddington closures correspond to solid and dashed curves. At the bottom we show the forces acting on the system, namely the radiative force and the fluid's gradient pressure.}
\end{figure}

This is Test 4$a$, with the characteristic that in the upstream zone, the ratio between radiation and gas pressures is $P_{r,L}/P_L\sim10$, larger than in Tests 1, 2, and $3a$,  with values $\sim 2\times 10^{-4}$, $\sim 1.6\times 10^{3}$, and $\sim 1.1\times 10^{-2}$, respectively. 

We show a snapshot of the results at $t = 3500$ in Figure \ref{fig:test4a} for the fluid density and velocity, radiation energy density, radiation flux and the forces due to radiation pressure and pressure gradient. Soon after the initial time, the configuration approaches a steady state when using the Eddington closure, whereas when using the M1 closure the solution drifts with a constant small speed $v_\text{sh} \sim 2.6\times 10^{-4}$ and in  the frame moving together with the drifting the configuration shows a steady solution.

Notice that the radiation flux $F^{x}_\text{r,co}$ is negative for the two cases on the left part of the domain, indicating the radiation energy is transferred from right to left. The radiation energy and pressure are dominant in this case, and the effects of radiation on the fluid affect the velocity, which shows a decreasing profile starting at $x\sim -10$ at the time of the snapshot. This is an effect understood as force produced by the radiation, which also  influences the increase of gas density  in the left part of the domain. In order to complete this interpretation, in the last plot of Figure \ref{fig:test4a} we show that the radiation force dominates over the pressure gradient.

	Test $4b$ is the optically thick version of Test $4a$. In Figure \ref{fig:test4b}, we show $\rho$, $v_x$, $E_\text{r,co}$, $F^{x}_\text{r,co}$, and the forces acting on the fluid  at $t = 3500$. Here we found that, the velocity/density begins to decrease/increase at  around $x \sim -2$. Also, the shock front travels across the domain with a drifting speed $v_\text{sh} \sim 2.55 \times 10^{-4}$ 
in code units, which is smaller than that of Test 4$a$. This is in concordance with the fact that the shock front in this case, propagates in a medium with a higher opacity than in Test 4$a$.
	
    These two tests verify that the code is able to resolve radiation-pressure dominated waves in two different optical depth regimes. \\
	
\subsection{Mildly-relativistic, optically thick flow}
	
\begin{figure}
\includegraphics[width=8.5cm]{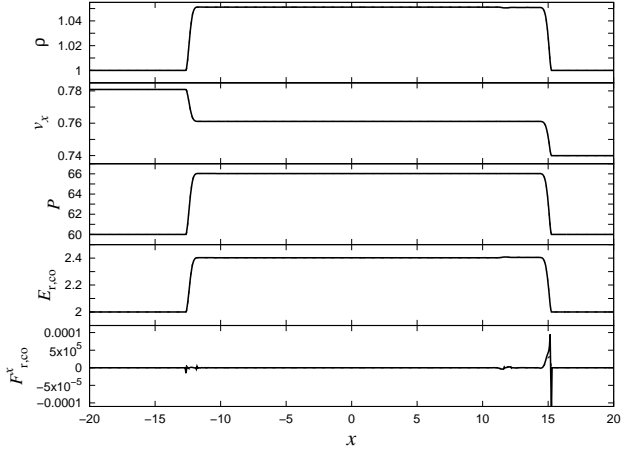}
\caption{\label{fig:test5b} 
Snapshot of the results for Test $5$ at time $t=15$. Shown are  the rest-mass density, velocity and pressure of the fluid, radiation energy density and  radiation flux. Results for the M1 and Eddington closures correspond to solid and dashed curves.}
\end{figure}
	
	This case is illustrated by Test 5 and is the only test that does not approach a stationary solution in the long term. The initial left and right states are identical except that they have different velocities. As a result, two shock waves propagate in opposite directions as shown during a snapshot in Figure \ref{fig:test5b}. This test is relevant because it represents an appropriate problem that tests the ability of a code to handle the stiffness of the source terms (\ref{eq:G0}) and (\ref{eq:Gi}). In this case, the thermal opacity coefficient is $1000$, much bigger than the one used in the previous tests. The results can be compared with those in \citep{Roedig_et_al_2012}, which confirms that {\it CAFE-R}  works satisfactorily in  optically thick and mildly-relativistic regimes.

\subsection{Radiative pulse}
\label{subsec:pulse}

In order to test the accuracy in an optically thick regime, we present a  radiative pulse similar to that described in \citep{Sadowski_et_al_2013b,McKinney_et_al_2014}, which consists in a radiative pulse propagating along one of the spatial cartesian coordinates in  Minkowski space-time.  This pulse has a profile given by  

\begin{equation}
  T_\text{rad}=\left(\frac{E_r}{a_r}\right)^{1/4}= T_0\left(1 + 100e^{(-x^2/\omega^2)}\right),	
\end{equation}

\noindent where $T_0=10^6$, $\omega=5$ and $a_r=6.24\time10^{-40}$. In this Test, unlike all the previous ones we assume that the thermal opacity coefficient is  zero $\kappa^t=0$, whereas the scattering opacity coefficient is  high $\kappa^s=10^{3}$. The background fluid is characterized by a constant rest-mass density and temperature given by $\rho=1$ and $T_\text{fluid}=T_0$ respectively. We solve this problem in the numerical domain $x \in[-50,50]$ with 200 cells and periodic boundary conditions. 

\begin{figure}
\centering
\includegraphics[width=8.9cm]{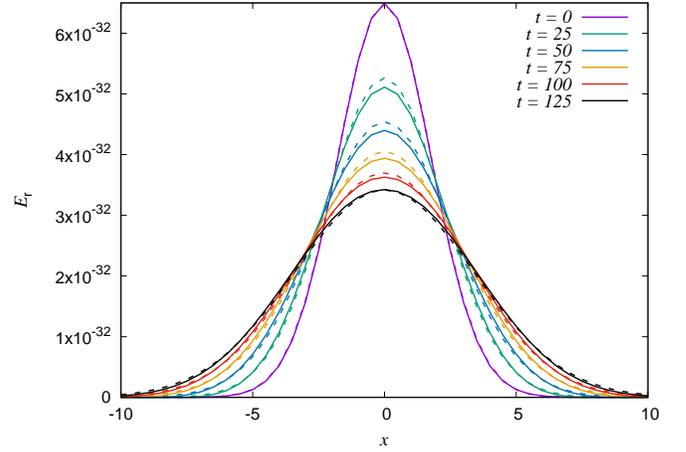}
\caption{\label{fig:thickpulse} Behavior of the radiative energy density in the laboratory frame for the pulse in the optically thick regime at different times. Solid lines show the numerical solution, while dashed lines represent the exact solution of Eq. (\ref{eq:diff}). The results are shown only in the domain $x\in[-10,10]$.}
\end{figure}

In this regime the dynamical evolution of the system can be described by the diffusion equation 

\begin{equation}
	\partial_tE_r=\frac{1}{3(\chi^t+\chi^s)}\partial_{xx}E_r,
	\label{eq:diff}
\end{equation}

\noindent whose analytical solution is $E_r(x,t) = 6.49\times10^{-32}\text{exp}\left(\frac{-x^2}{4t/3\chi^s}\right)\left(\frac{4\pi t}{3\chi^s}\right)^{-1/2}$. In Figure \ref{fig:thickpulse}, we show the numerical solution (solid lines) for the radiative energy density compared with the exact solution of Eq. (\ref{eq:diff}) (dashed lines) at various instants. We can see that the numerical solution diffuses slightly faster than the analytical solution, which is due to the additional numerical dissipation introduced by the the numerical scheme. However, at later times this difference becomes small. Our numerical solution is in agreement with that obtained by \citep{Sadowski_et_al_2013b,McKinney_et_al_2014}. 

\subsection{Single beam}

After showing the code works fine in 1D problems, we start with 2D tests. In order to carry out these tests with the 3D driver, we use 5 cells along the additional direction and impose outflow boundary conditions along this additional direction.

First we present the single beam problem, a test intended to show the  capability to simulate a scenario where the fluid and radiation are decoupled. These conditions corresponds to an optically thin medium with zero opacity $\kappa_a=\kappa_{total}=0$. This specific problem consists in the evolution of a beam and check it does not break or distorts during the evolution.

We simulate the process in the domain $[0,1]\times [0,1]$ using $50\times50$ grid cells. The beam is constantly injected in the segment of boundary given by $x=0$, $y\in[0.4,0.6]$ and the rest of the boundary uses outflow conditions. For this exercise the injected radiation energy density is $E_{r,beam}=10^7$, whereas in the rest of the domain $E_r=10^5$. We use $a_r =Er/(P/\rho)^4 \sim 10^{17} $ and set $\Gamma =4/3$. The hydrodynamical variables remain constant throughout the domain with values $\rho=1$, $P=10^{-3}$ and $W=1$.

\begin{figure}
\centering
\includegraphics[width=8.9cm]{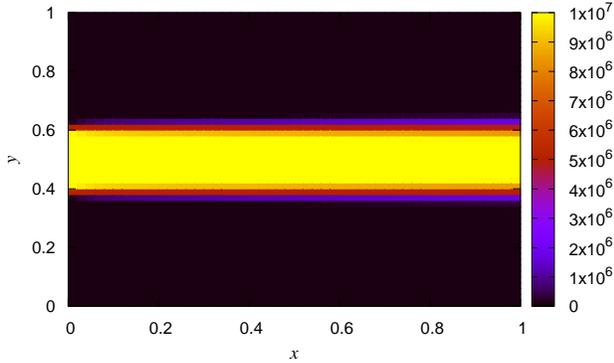}
\caption{\label{fig:SingleBeam} Radiation energy density at $t=10$. The radiation beam is being injected from the left boundary.}
\end{figure}

A snapshot of $E_r$ is shown in Fig. \ref{fig:SingleBeam} at $t=10$ and shows that the beam propagates through with no distortion nor disruption. The  result is equivalent to that in \cite{Sadowski_et_al_2013a} and \cite{takahashi_Ohsuga_2013}.

\subsection{Pulse collision}

In this test, two pulses of radiation propagate along the diagonal directions of the domain and eventually interact.
This test is solved in the  domain $[0,1]\times [0,1]$ at the plane $z=0$, that we cover with $100\times100$ grid cells. The pulses are injected from the two regions defined by the boundary segments $x=0$ and $x=1$ for  $y=[-0.875,-0.750]$. We use outflow boundary conditions at the  boundary, except at the segments where the beams are  injected. The radiative flux components  are $F^i_r= (E_r/\sqrt{2},E_r/\sqrt{2},0)$, where the value of the radiation energy density is $E_r=20$ and the radiation constant $a_r=2\times 10^{-7}$. Both pulses propagate through a static background, this means that the hydrodynamical variables are kept fixed during the evolution and only the radiation variables evolve. In this test  we use $\rho = 5$ and adiabatic index $\Gamma = 5/3$.

\begin{figure}
\centering
\includegraphics[width=8.9cm]{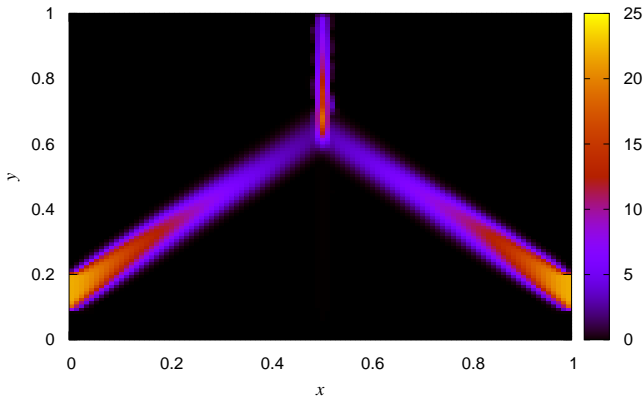}
\caption{\label{fig:PulseCollision} Radiation energy density at $t=10$. The radiation beams are located at segments of the left and right boundaries and are launched along lines with slope of $\pm$45 degrees.}
\end{figure}

In Figure \ref{fig:PulseCollision} we show a snapshot of the radiation energy density in the laboratory frame at $t=10$. We can see that the beams propagate in straight directions until they interact. The interaction is such that the two beams do not pass through each other, instead  they produce a beam in the $y-$direction. This is consistent with the fact that the radiation flux component $F^x_r$ becomes zero when both pulses interact. In the formulation of the M1-closure, the radiation stress tensor $P_r^{ij}$ is determined only by $E_r$ and $F^i_r$, but it is independent of the optical depth. This means that the closure relation approaches that of the Eddington approximation when the flux vanishes even when the system is optically thin. This is an apparent inconsistency of the M1-closure model, but it is not, this happens because the M1-closure scheme can distinguish only a single direction of radiative flux \citep{Levermore_1984}. In order to resolve this problem, we need to implement a more general relation closure, for instance by using the variable Eddington tensor scheme, which handles multiple directions of the radiative flux at a single numerical cell \citep{Gehmeyr_Mihalas_1993,Stone_et_al_1992,Hayes_Norman_2003,Jiang_et_al_2012,Ohsuga_Takahashi_2016}.

\subsection{Shadow}

This problem serves to verify that the M1 approximation works properly, and to illustrate the difference between  M1 and Eddington approximations in the presence of an obstacle. 
The test consists in the injection of the gas-radiation fluid on an optically thin environment that encounters a circular obstacle made of  an optically thick gas. We solve the test on the domain $[-1,3]\times[-1,1]$ on the plane $z=0$, that we cover with $100\times50$ cells. Following the set up in \citep{takahashi_Ohsuga_2013}, the mass density within the circle is given by

\begin{equation}
    \rho_0 = \rho_a + (\rho_b - \rho_a)e^{(-\sqrt{x^2+y^2}/\omega^2)},
\end{equation}

\noindent where $\rho_a=10^{-4}$ and  $\rho_b=10^3$ are the ambient and injected densities, whereas $\omega=0.22$ defines the size of the circular distribution acting as an obstacle. In this test, not only radiation is being injected in this test, but matter as well. At initial time the fluid is at rest, the fluxes are set to zero and thermal equilibrium is assumed.

The fluid temperature is estimated considering its pressure is constant throughout the domain

\begin{equation}
 	T_\text{fluid} = T_a\frac{\rho_a}{\rho_0}.
\end{equation}

\noindent where $T_a$ is the initial temperature of the system. We assume the beam enters the domain from the left side of the domain and leaves through the right side, for which we assume inflow and outflow boundary conditions at these faces respectively, whereas we set periodic conditions in the other faces. The properties of the beam are as follows $E_L=a_rT^4_{\text{fluid},L}$, $F^x=0.99999E_L$, and $T_{\text{fluid},L}=100T_a$, for which we have assumed $T_a=10^{10}$, which is four orders of magnitude bigger than in \cite{Sadowski_et_al_2013b}, only to show a test with different parameters.
Following \cite{Sadowski_et_al_2013b}, in these tests we set the radiation constant to $a_r=351.37$, the opacities $\kappa_a=\kappa_{total}=\rho_0$ and the adiabatic index to $\Gamma=1.4$.

\begin{figure}
\centering
\includegraphics[width=8.9cm]{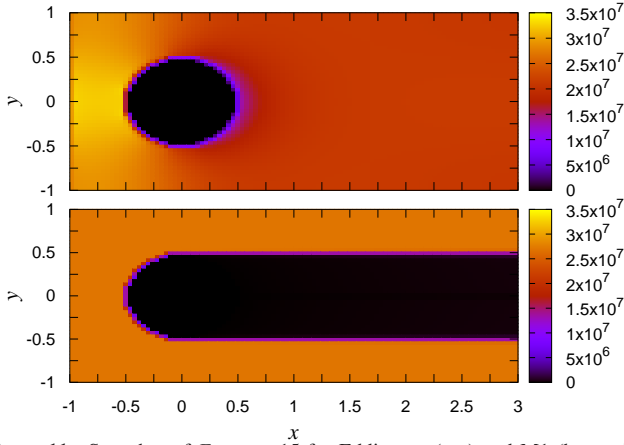}
\caption{\label{fig:Shadow} Snapshot of $E_r$ at $t=15$ for Eddington (top) and M1 (bottom) closure models.}
\end{figure}

A snapshot of $E_r$ at $t\sim15$ is shown in Figure \ref{fig:Shadow} for the two closure models. As expected for the Eddington closure, the radiation field  propagates isotropically, including the region behind the obstacle and no shadow is expected to form. On the other hand, the solution with the M1 approximation corresponds to a radiation field propagating parallel to the direction used for the injection in this optically thin medium where $F_r\approx E_r$, producing the shadow that can be compared with that in \cite{Sadowski_et_al_2013b}.

\subsection{Double shadow}

This tests shows the capability to deal with scenarios with various light sources and for which we follow the standard set up in \cite{Sadowski_et_al_2013b}, except that we again use a temperature four orders of magnitude bigger $T_0=10^{10}$. We set this test in the domain $x\in[-6,3] \times y\in[-1.5,1.5]$ with a circular obstacle at the origin with the same parameters as in the single shadow test.

The initial conditions for fluid and radiation are exactly the same as for the simple shadow test, but unlike that test, the beam is injected differently from the left boundary as follows. For $y>0.3~(y<-0.3)$ at $x=-6$, the horizontal and vertical radiation flux components are $F^x_r=0.93E_r~(F^x_r=0.93E_r)$, $F^y_r=-0.37E_r~ (F^y_r=0.37E_r)$, which models two light beams, a first one launched from the upper side of the boundary moving downwards and a second one launched from the lower part of the boundary moving upwards. 

\begin{figure}
\centering
\includegraphics[width=8.9cm]{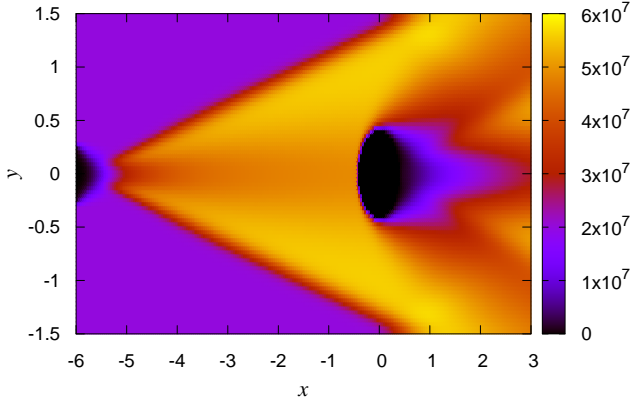}
\caption{\label{fig:DoubleShadow} Snapshot of $E_r$ at $t=20$ for the Double Shadow test.}
\end{figure}

Eventually the light beams interact, a process we simulate using the M1 closure model and a snapshot of $E_r$ showing this interaction at $t=20$ is shown in Figure \ref{fig:DoubleShadow}. At the center of the left boundary there is a spot  where the radiation does not enter, later on there are two  straight lines where $E_r$ changes from a low to a high value, where the two beams superpose. More to the right, there is the region past the obstacle which has the umbra with low $E_r$ and the penumbra which is a zone of partial shadow. Even though the M1 closure is used, there is an unphysical low radiation energy zone that should not be there at the $x-$axis to the right from the obstacle. This shows that even though the M1 seems to produce more consistent  results compared with the Eddington model, it has weaknesses in systems with multiple sources. Finally, considering this limitation of the M1 closure, the results in the  snapshot of Fig. \ref{fig:DoubleShadow} are consistent with those in \cite{Sadowski_et_al_2013b}.

\subsection{Radiation pulse in 3D}

\begin{figure}
\centering
\includegraphics[width=8.9cm]{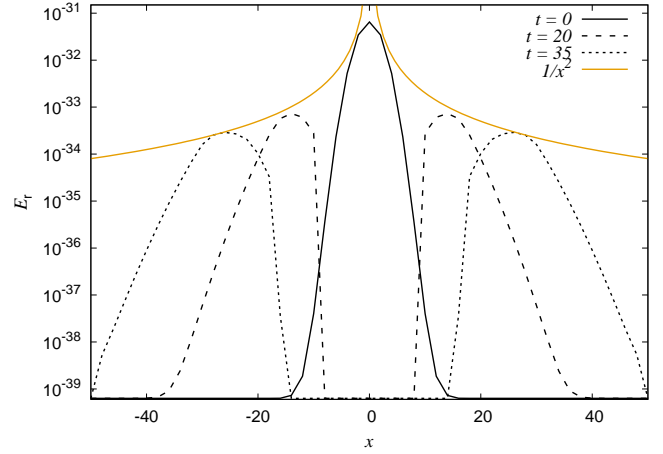}
\caption{\label{fig:thinpulse} Profiles of the energy density along the $x-$direction at different times, such profiles follow the expected trend of energy decrease from the center of the domain with a power law $\sim 1/x^2$ (orange line).}
\end{figure}

Finally, there is a second test related to a pulse, similar to that in Section \ref{subsec:pulse} but this time in 3D within the optically thin regime. It consists in the evolution of  a Gaussian distribution of radiative energy density  centered at the origin. In this case, the radiative temperature is set to  \citep{Sadowski_et_al_2013b,McKinney_et_al_2014}

\begin{equation}
  T_\text{rad}=\left(\frac{E_r}{a_r}\right)^{1/4}= T_0\left(1 + 100e^{(-(x^2 + y^2 + z^2)/\omega^2)}\right),
\end{equation}

\noindent with $\kappa^t=\kappa^s=0$. We solve this problem in the domain $[-50,50]^3$ covered with with $50\times50\times50$ cells. In  Figure \ref{fig:thinpulse} we show the evolution of the radiative energy density. Since the background is an optically thin medium, the initial pulse is expected to spread isotropically with velocity close to the speed of light. Also, as a consequence of the energy conservation, the energy density decreases as one over the square of the distance to the origin. We show this behavior in Figure \ref{fig:thinpulse}, where  $E_r$ decreases  with distance from the center with approximately the appropriate power law.

\subsection{Self-convergence}
\label{sec:resolution}

We now present  self-convergence for the one dimensional tests. For a given variable $\phi$, that could be any of the primitive variables of the equations, assuming there is an exact solution $\phi^e$, 
one calculates three numerical solutions  $\phi_1$, $\phi_2$ and $\phi_3$ using the resolutions $\Delta x_1$, $\Delta x_2$ and $\Delta x_3$, with $\Delta x_3 < \Delta x_2 < \Delta x_1$. If our methods are accurate to order $n$, then at each point of the domain

\begin{eqnarray}
\phi_1 &=& \phi^e + E \Delta x_1 {}^{n},\nonumber\\
\phi_2 &=& \phi^e + E \Delta x_2 {}^{n}, \nonumber\\
\phi_3 &=& \phi^e + E \Delta x_3 {}^{n}, \nonumber
\end{eqnarray}

\noindent where $E$ is the amplitude of the error at a given point of the numerical domain at a given time slice. By combining the above equations one has the following comparison among the numerical solutions that define the self-convergence factor to be

\begin{equation}
CF = \frac{\phi_1 - \phi_2}{\phi_2 - \phi_3} = \frac{\Delta x_1^n - \Delta x_2^n}{\Delta x_2^n - \Delta x_3 ^n},
\label{eq:cf}
\end{equation}

\noindent a relationship that is expected to be approximately fulfilled at every point of the domain and at every snapshot of the evolution. For instance in regions where the fields are smooth, the methods used in $CAFE-R$ explained above, are accurate to order $n=2$ whereas in regions with discontinuities the accuracy is of order $n=1$. In general the solution is polluted  with the poorest accuracy $n=1$ and CF varies between the value corresponding to $n=1$ and the value corresponding to $n=2$ in (\ref{eq:cf}).

Since it is impractical to calculate the self-convergence across the whole numerical domain over the whole evolution, it is common to test the self-convergence at a given time using a norm of $\phi_1 - \phi_2$ and $\phi_2 - \phi_3$. In our case, we calculate the self-convergence at an interesting snapshot, specifically at $t=3500$ as shown in Figures from 1 to 7, using the $L_1$ norm of the difference between numerical solutions. The convergence factor is then $CF=||\phi_1 - \phi_2||_1/||\phi_2 - \phi_3||_1 $, for all the five tests given in  Table \ref{table:testGB}. For the first, second, and third resolutions we use $800$, $1600$ and $3200$ cells to cover the  domain. Specifically, this corresponds to resolutions $\Delta x_1=0.05$, $\Delta x_2=0.025$, and $\Delta x_3=0.0125$. 

In Table \ref{table:CF}, we report the self-convergence factors for the rest-mass density and the radiation energy density of each test at two different times. We point out that the stationary stage is achieved at around $t\sim 1000$ for all tests, and in the Table we show the self convergence factor is kept within first ($CF=2$) and second ($CF=4$) order of convergence until $t=3500$, which is consistent with the theory in (\ref{eq:cf}). In the Table we also show how the self-convergence degrades by time $t=5000$ down to below first order  ($CF \lesssim  2$).

	\begin{table}[h!]	
	\begin{center}
	\begin{tabular}{c  c  c| c c }
	\hline
	  $t=3500$   & \multicolumn{2} {c|} {Eddington } & \multicolumn{2} {c} {M1 } \\
	\hline
	Test & CF($\rho$) & CF($E_\text{r}$)& CF($\rho$) & CF($E_\text{r}$) \\
	\hline
	 $1$ &   2.6  &    2.3   & 2.6 &  2.52  \\
	\hline
	 $2$ &   2.29 &    2.32   &  2.51 & 3.4  \\
	\hline
	 $3a$&   2.42 &   2.35   &  2.23  &  2.22 \\
     \hline
     $3b$&   3.24 &    3.41   &  3.32   &  3.51  \\
	 \hline
	 $4a$&   2.36 &    2.35  & 3.8 & 3.2   \\
     \hline
     $4b$&   2.21 &    2.33  & 2.32 & 2.35  \\
	\hline
	\hline
	  $t=5000$   & \multicolumn{2} {c|} {Eddington } & \multicolumn{2} {c} {M1 } \\
	\hline
	Test & CF($\rho$) & CF($E_\text{r}$)& CF($\rho$) & CF($E_\text{r}$) \\
	\hline
	 $1$ & 0.9        &       2.2    &      0.83   &   1.9 \\
	\hline
	 $2$ & 0.8        &   1.99       &    0.6      & 0.9 \\
	\hline
	 $3a$& 1.35       &   1.33       &   0.8      & 0.76    \\
     \hline
     $3b$& 3.25       & 3.28         & 3.1        &  3.42  \\
	 \hline
	 $4a$& 2.231      & 2.23         & 1.9        &  1.7   \\
     \hline
     $4b$&  1.75      &  1.98        &  1.98      & 2.004 \\
     \hline\hline
	  $t=7$   & \multicolumn{2} {c|} {Eddington } & \multicolumn{2} {c} {M1 } \\\hline
	Test & &&&\\
     $5$ &  2.041 &   2.006  & 2.068 & 2.011 \\\hline\hline
	  $t=15$   & \multicolumn{2} {c|} {Eddington } & \multicolumn{2} {c} {M1 } \\\hline
	Test & &&&\\ 
     $5$ &  2.01 &   2.001  & 2.067 & 2.02 \\\hline
	\end{tabular}
	\caption{\label{table:CF} Self-convergence factor for the rest-mass density $\rho$ and radiation energy density $E_r$ in the 1D shock tube tests using Eddington and M1 closures. For tests from 1 to 4 the snapshots for convergence are chosen at $t= 3500 ~\text{and}~ 5000$ whereas for test 5 these are taken at $t=7~\text{and}~15$.}
	\end{center}
	\end{table}

\section{Jets}
\label{sec:jets}

One of the most energetic events in astrophysics are the jets, which are produced by the death of supermassive stars, or active galactic nuclei (AGNs). Emulating these events using  numerical simulations  is challenging and a test of stability, in particular, because of the formation of strong external and internal shocks.  This is the reason why, as an illustrative application, in order to test the potential of using {\it CAFE-R} in 3D, we present the simulation of a non-axisymmetric jet. 

For this we assume the jet is produced by the injection of a relativistic beam  from a nozzle with radius $r_\text{b}$ and velocity $v_\text{b}$. This model is characterized by the ratio between the density of the beam (subindex $b$) and that of the medium (subindex $m$)  $\eta=\rho_\text{b}/\rho_\text{m}$ and by the ratio between their pressures $K = P_\text{b}/P_\text{m} $. The relativistic Mach number of the beam is $M_\text{b}={\cal M}_\text{b}W_\text{b}\sqrt{1-c_\text{s}^2}$, where ${\cal M}_\text{b}$, $W_\text{b}$, and $c_\text{s}^2$ are the Newtonian Mach number, Lorentz factor, and speed of sound, respectively. The boundary conditions  we use are outflow in all borders, except at the nozzle located at the center of one of the boundary faces, where the values of the variables are kept constant in time during the injection.
 
For comparison, we analyze two cases: (i) a purely hydrodynamical jet (HD-jet) and (ii) a radiation-pressure-dominated jet (RRH-jet). 
We perform this simulation on the domain $[-7.5,7.5]\times[-7.5,7.5]\times[0,75]$ with resolution  $\Delta x=\Delta y=\Delta z=0.125$.
The jet is injected at the plane $z = 0$ toward the positive $z-$axis through a circular nozzle of the radius $r_\text{b}=1$. 
The resolution and size of the nozzle are such that the later contains at least eight cells per beam radius, which is a recommended resolution to properly resolve the internal structure of the jet and its interaction with the external medium \citep{Aloy_et_al_1999,Aloy_et_al_2000}. Moreover, we choose the density and pressure ratios to be initially $\eta=0.01$ and $K = 1$, the beam Lorentz factor $W_\text{b} = 7$, the Mach number $M_\text{b}=6$ and the adiabatic index $\Gamma=5/3$. 

For the RRH-jet, the radiative energy density is chosen such that the radiation pressure is dominant over the gas pressure. For this, we set $E_\text{r,b} = 10^{-3}$, which gives a pressure ratio of $\frac{P_\text{r,b}}{P_\text{b}}= \frac{E_\text{r,b}}{3 P_\text{b}}\sim 2$. The opacity coefficients are constant in space and time given by $\kappa^t = 1$ and $\kappa^s = 10^{-3}$. Thus, 
since the system is in an optically thick regime, we can choose a small value of the radiative flux of the beam, such as  $F_\text{r,b} = 10^{-2}E_\text{r,b}$. 

The 3D nature is imposed by a helical perturbation in the velocity and radiative flux profiles at the nozzle, following the implementation of helical perturbations on hydrodynamical jets in the past \citep{Aloy_et_al_1999}

\begin{eqnarray}
v^x_\text{b}   &=& \varsigma v_\text{b}\cos\left( \frac{2\pi t}{\tau}\right), \nonumber\\
v^y_\text{b}   &=& \varsigma v_\text{b}\sin\left( \frac{2\pi t}{\tau}\right), \nonumber\\
v^z_\text{b}   &=& v_\text{b}\sqrt{1 - \varsigma^2}, \nonumber\\
F^x_\text{r,b} &=& \varsigma_\text{r} F_\text{r,b}\cos\left( \frac{2\pi t}{\tau}\right), \nonumber\\
F^y_\text{r,b} &=& \varsigma_\text{r} F_\text{r,b}\sin\left( \frac{2\pi t}{\tau}\right), \nonumber\\
F^z_\text{r,b} &=& F_\text{r,b}\sqrt{1 - \varsigma_\text{r}^2}, \nonumber
\label{eq:velJet}	
\end{eqnarray}
 
\noindent where $\varsigma=0.01$ and $\varsigma_\text{r}=0.01$ are perturbations of the helical velocity and helical radiative flux, respectively and $\varsigma_\text{r}=0$ for the HD-jet. The quantity $\tau=T/n$ is the perturbation period, where $n=200$ is the number of the cycles completed during the  injection and $T\sim 150$ is the injection time of the jet.

\begin{figure*}
	\centering
\includegraphics[width=0.33\textwidth,height=0.55\textheight]{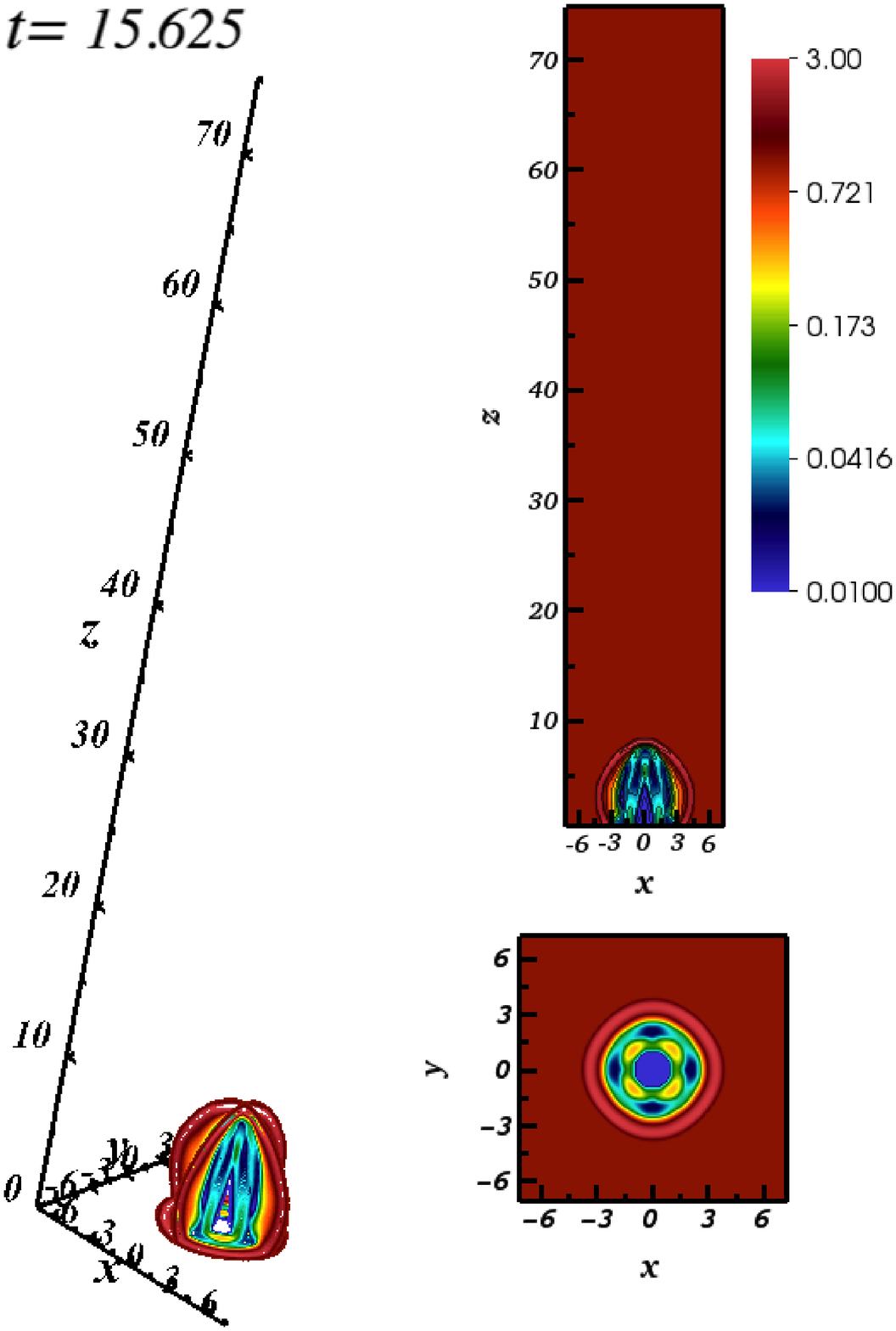}
\includegraphics[width=0.33\textwidth,height=0.55\textheight]{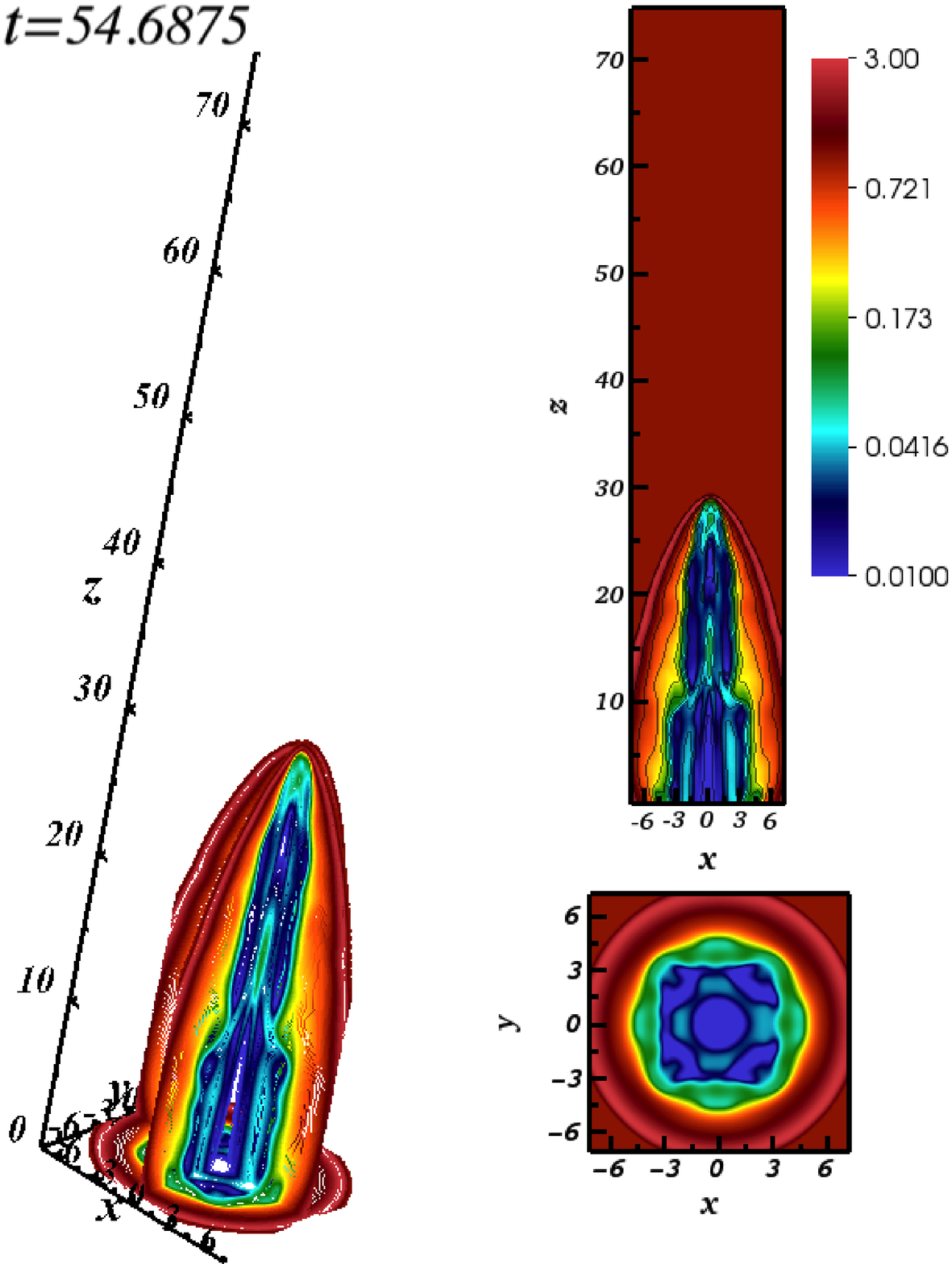}
\includegraphics[width=0.33\textwidth,height=0.55\textheight]{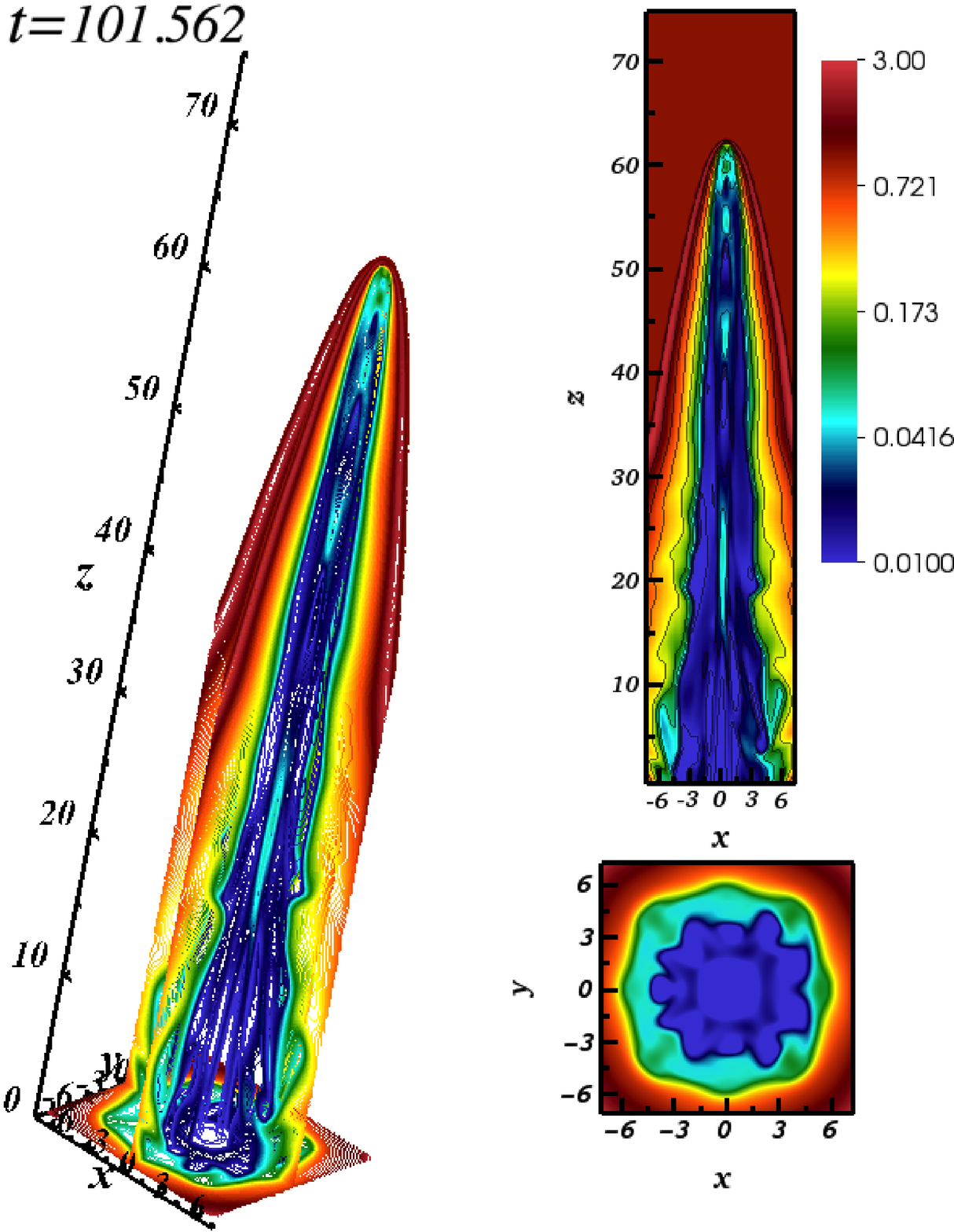}
\caption{\label{fig:PurelyHydro} Snapshots at different times of the rest-mass density iso-surfaces, and at the planes $y=0$ and $z=0$ for the HD-jet.}
\end{figure*}

In Fig. \ref{fig:PurelyHydro}, we show the behavior of the purely hydrodynamical jet at different times. We can see snapshots of the rest-mass density, viewed from different perspectives. We show 3D snapshots of the contours, a slice of the plane $y=0$ and a slice at plane $z=0$. At $t\sim 15.6$, we can see the basic characteristics of the jet, namely a collimation shock in the beam, a bow shock,  the reverse shock produced by the interaction between the jet and the external medium, and the formation of a cocoon. At this time, we can see that the beam does not exhibit any twisting perturbation yet. Later on, at $t\sim 54.6$, we can see structures behind the head of the jet similar to Kelvin-Helmholtz instabilities, which could be better resolved using a higher resolution. Also, at this time, we can see that the helical perturbations exhibited by the beam are still unnoticed.  Finally, at $t\sim 101.5$, we can see how the helical perturbation in the velocity profile produces an asymmetric jet, as shown for HD-jets in the past \citep{Aloy_et_al_1999}.

\begin{figure*}
	\centering
\includegraphics[width=0.33\textwidth,height=0.55\textheight]{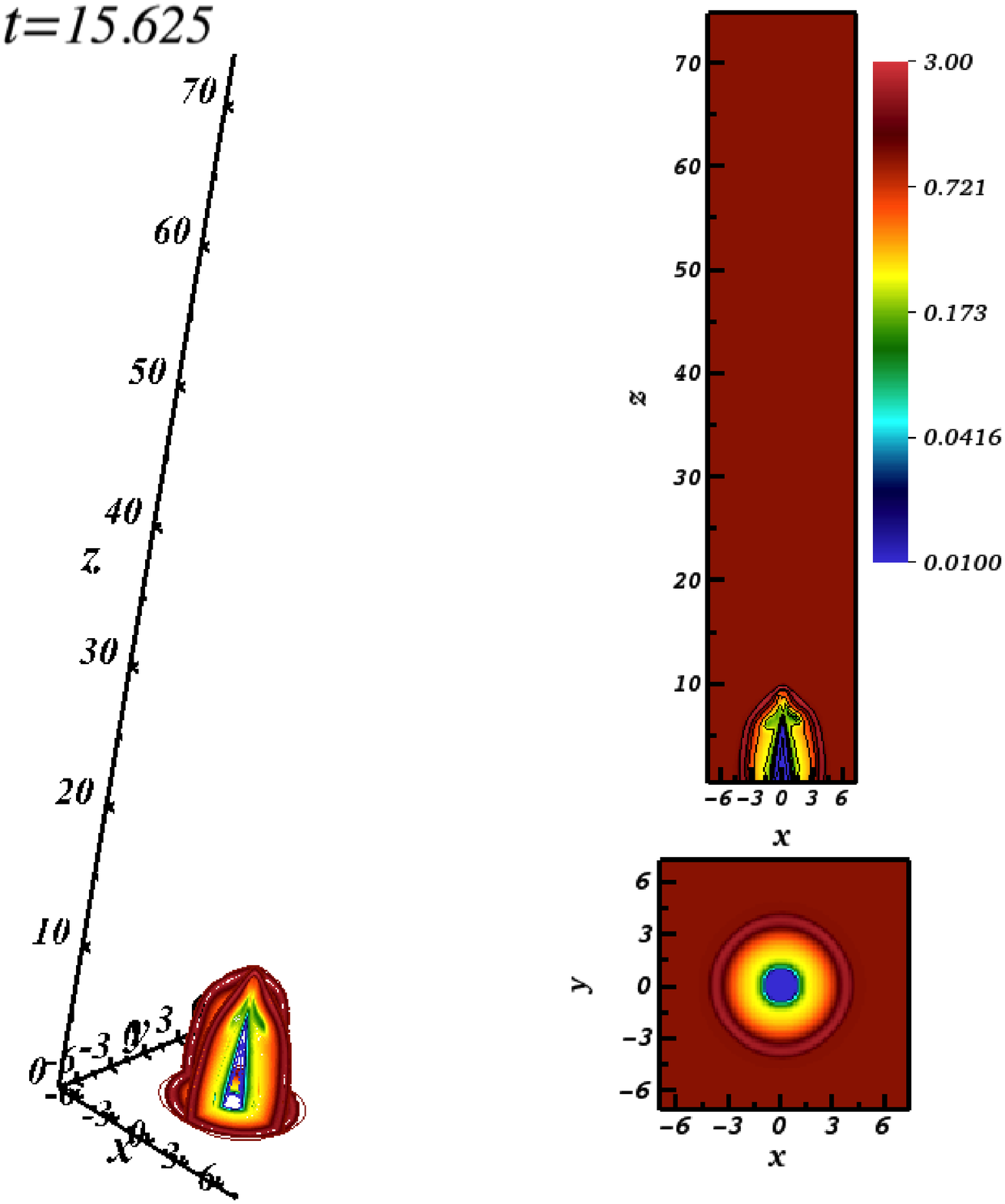}
\includegraphics[width=0.33\textwidth,height=0.55\textheight]{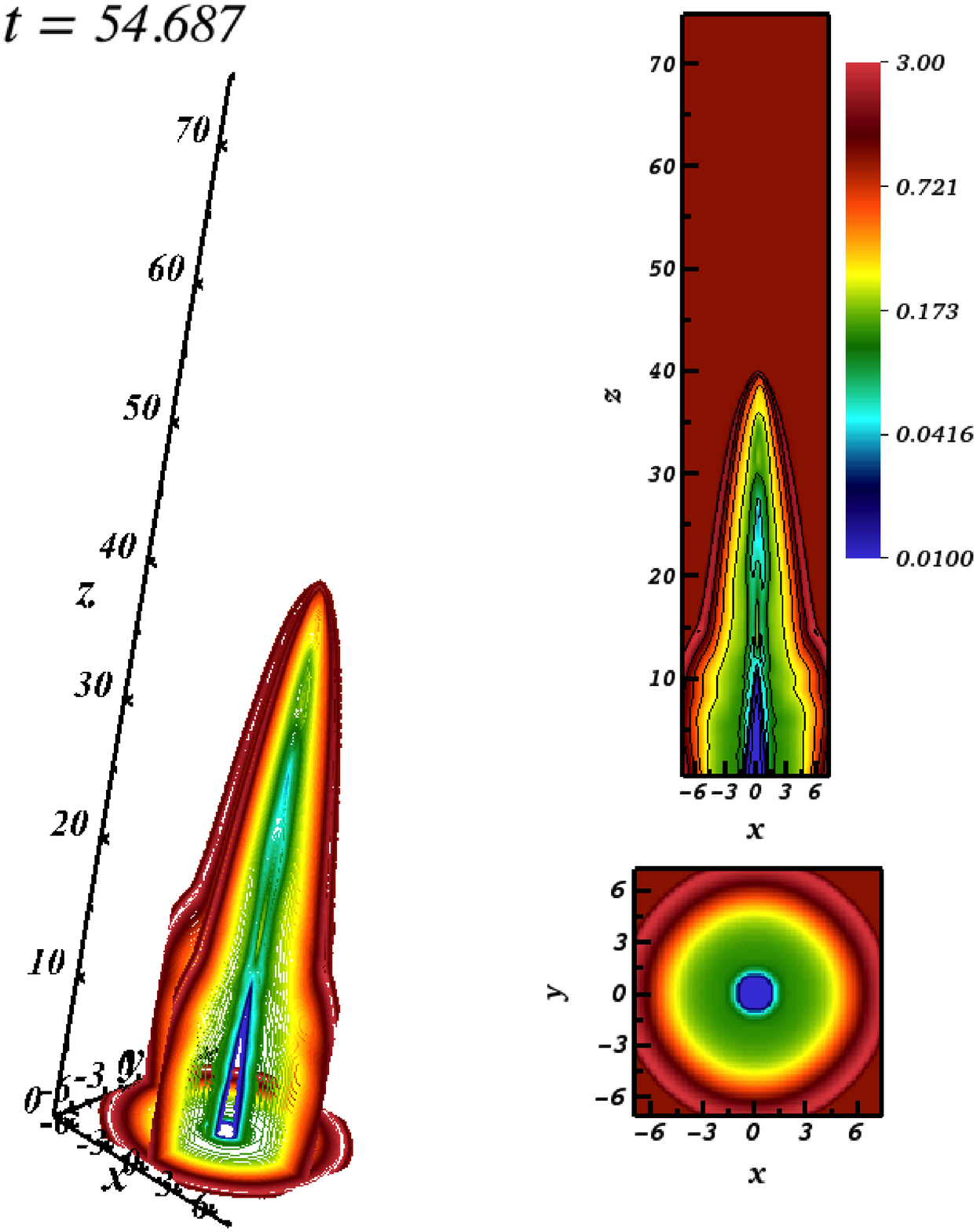}
\includegraphics[width=0.33\textwidth,height=0.55\textheight]{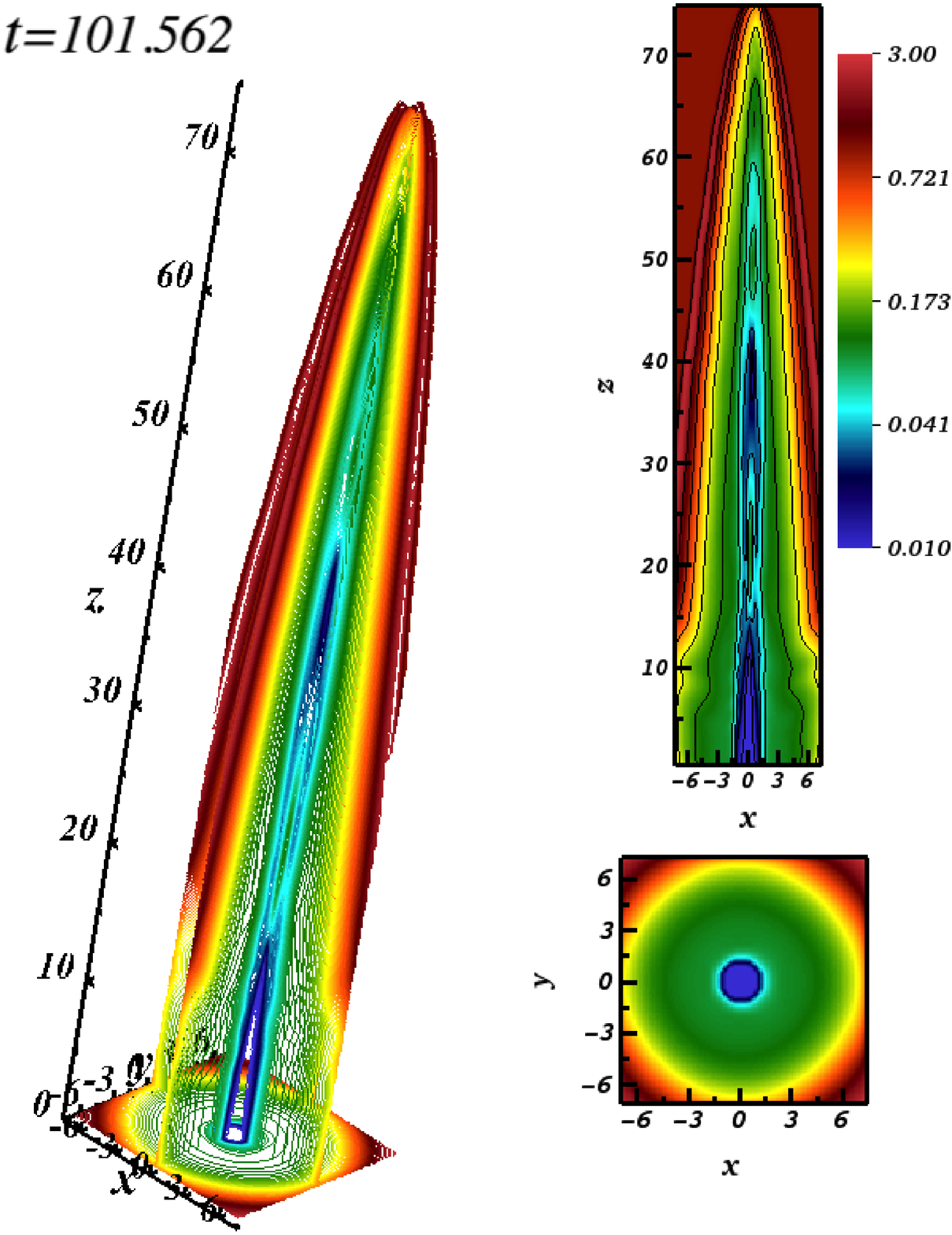}
\caption{\label{fig:RadPressDom} Snapshots at different times of the rest-mass density iso-surfaces, and at the planes $y=0$ and $z=0$ for the RRH-jet.}
\end{figure*}

In Fig. \ref{fig:RadPressDom}, we show the dynamical evolution of the RRH-jet at different times.  By $t\sim 15.6$, we can see how the high pressure at the cocoon compacts the jet and some material is convected backwards as well as a turbulent behavior behind the working surface. At this time, there is a big difference with respect to its HD-jet counterpart (Fig. \ref{fig:PurelyHydro}), in which the density does not show signs of a twisted profile, whereas in the RRH-jet it does, which can be seen  behind the jet's head. In both cases there is a  reverse shock from the contact discontinuity which modifies the structure of the jet head and influences the further propagation into the surrounding medium \citep{Massaglia_et_al_1996}. 
In this context, the radiation pressure helps to push material towards the contact discontinuity by making the internal shock of the RRH-jet behind the jet head stronger than in the HD-jet. At $t\sim 54.6$, the head jet keeps  spreading continuously upwards, which generates the expected  strong shock with the ambient medium. At this moment, the effect of the radiation pressure is noticeable, because it accelerates the matter faster than in the HD-jet case. Finally, at $t\sim 101.5$, radiation pressure pushes the matter off the domain.
 
It would be interesting to know whether the conditions in our simulations can trigger instabilities able to destroy the jet, which have been found in multidimensional jet evolutions as described in \citep{Matsumoto_Masada_2013,Gourgouliatos_Komissarov_2018}. According to \cite{Matsumoto_Masada_2013}, some radial oscillations can trigger Rayleigh-Taylor (RTI) and Richtmeier-Meshkov (RMI) instabilities. There is a  criterion to know whether the jet can be destroyed by those instabilities or not, indicating that the instabilities will be triggered if the inertia ratio between beam jet and the external medium  $\eta$ is less than the unity, where 

\begin{equation}
\eta = \frac{W^2_\text{b}(\rho_\text{b} h_\text{b}+4/3 E_\text{r,b})}{\rho_\text{m} h_\text{m}+4/3 E_\text{r,m}},	
\end{equation}

\noindent where $W^2_\text{b}\rho_\text{b} h_\text{b}$ is the effective inertia in the purely hydrodynamic case and the effective inertia of the radiation field is $W^2_\text{b} 4/3 E_\text{r,b}$. For both, HD-jet and RRH-jet, the initial effective inertia ratio is $\eta_\text{HD}\simeq 0.51$ and $\eta_\text{RRH}\simeq 0.558$ respectively. That means that both, RTI and RMI instabilities are suppressed and hence, the structure of the jets is expected to be kept  during the  evolution.

Among the differences between the two jets, there is a particularly interesting one related to  oscillations separated by about $\sim 10$ length units, of the layer between the jet and the ambient medium that can be seen in the $y=0$ plane at $t\sim 101$ for the HD-jet in Fig. \ref{fig:PurelyHydro}, that are not seen in the RRH-jet. A possible explanation is that the radiation smoothens out the oscillations, because the radiation diffusion length given by $L_\text{diff}=\sqrt{D T}$, where $D=1/(3\rho(\kappa^t+\kappa^s))$ (see e.g. \cite{Turner_Stone_2001}), for the parameters used in the RRH-jet is $L_\text{diff}\sim 70$, a few times bigger than the distance between oscillations.

\begin{figure}
	\centering
\includegraphics[width=0.4\textwidth]{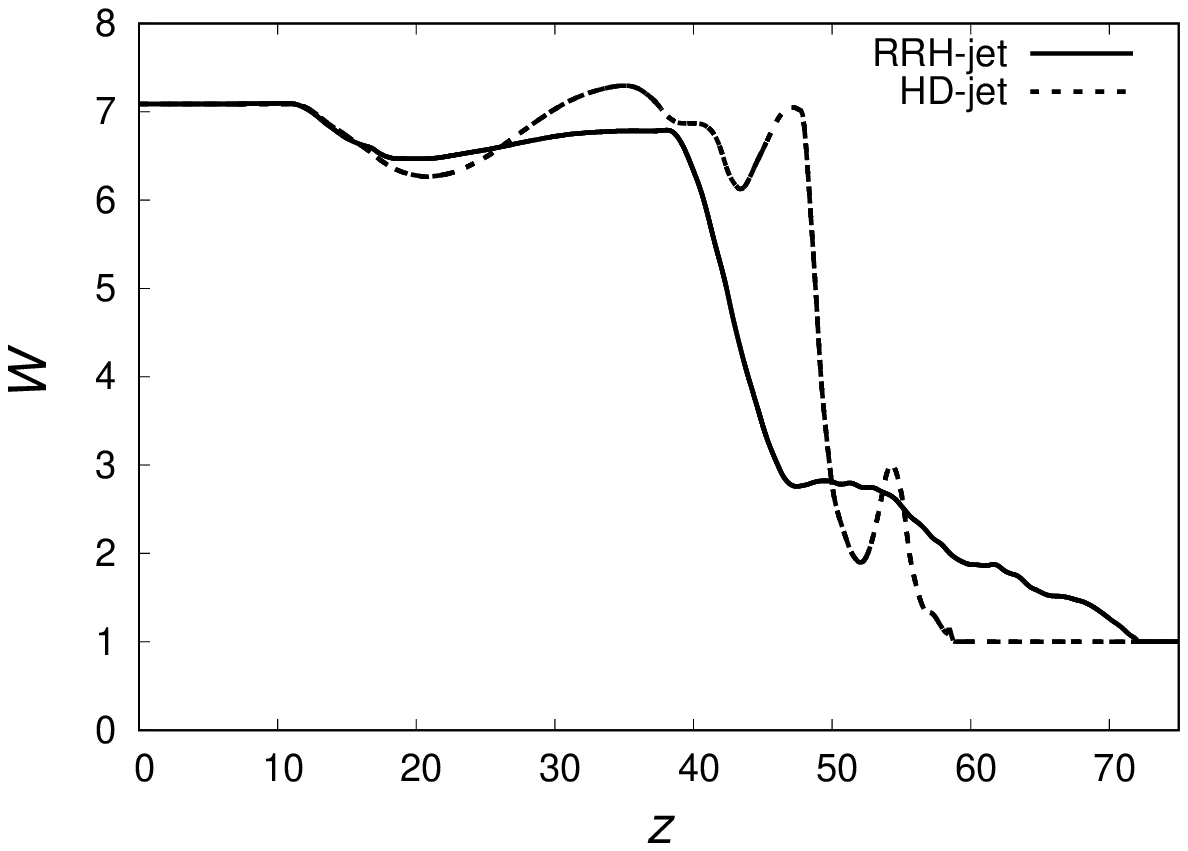}
\includegraphics[width=0.4\textwidth]{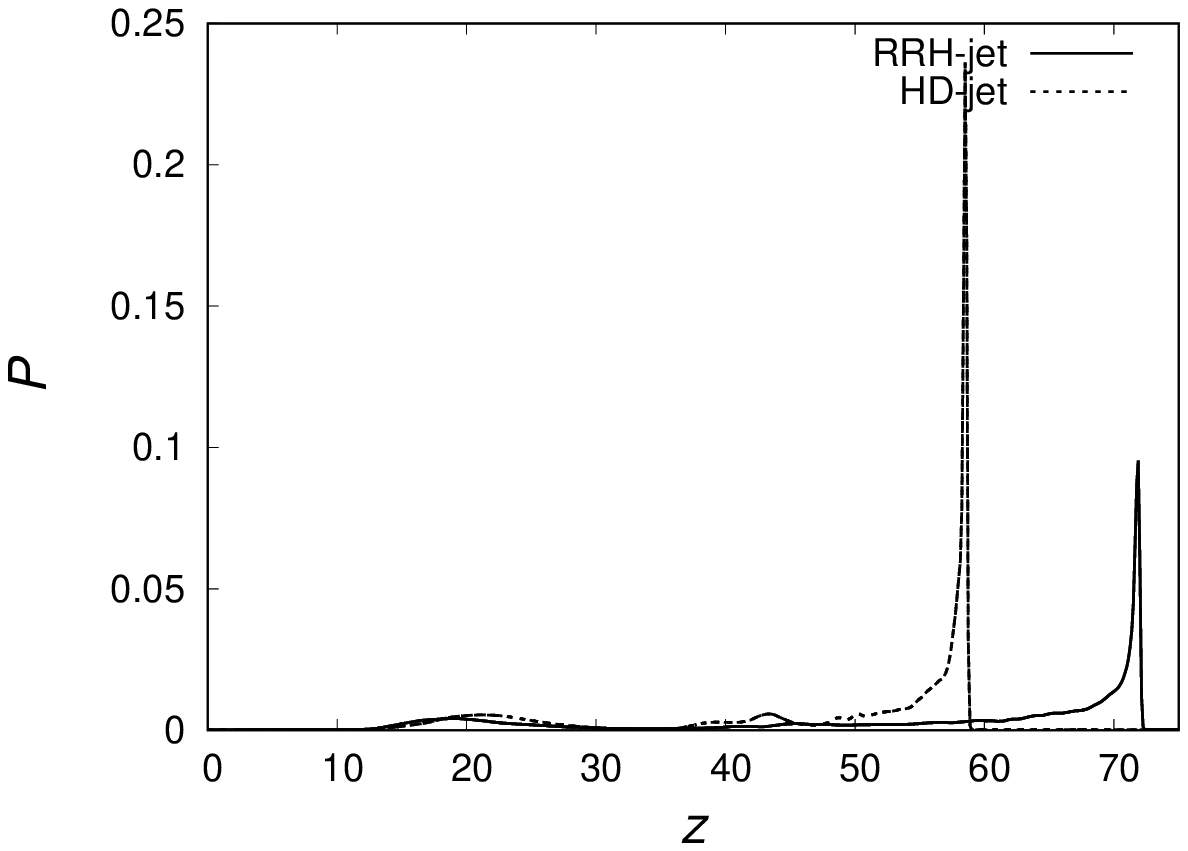}
\caption{\label{fig:comparation1} Lorentz factor and pressure along the $z-$axis for the HD and RRH jets at $t = 100$. The front shock moves faster in the RRH-jet than in the HD-jet case.}
\end{figure}

In order to describe the effect of  radiation, we compare the profiles of the pressure, rest-mass density, Lorentz factor, and the maximum of the temperature's fluid of the HD and RRH-jets, measured along the $z-$axis at the same time $t=100$. In Fig. \ref{fig:comparation1} we show the Lorentz factor, which  illustrates the influence of radiation pressure in the velocity of the jet. In the RRH-jet case, the pressure is lower and the radiation boosts the jet making it faster than the HD-jet.

\begin{figure}
	\centering
\includegraphics[width=0.4\textwidth]{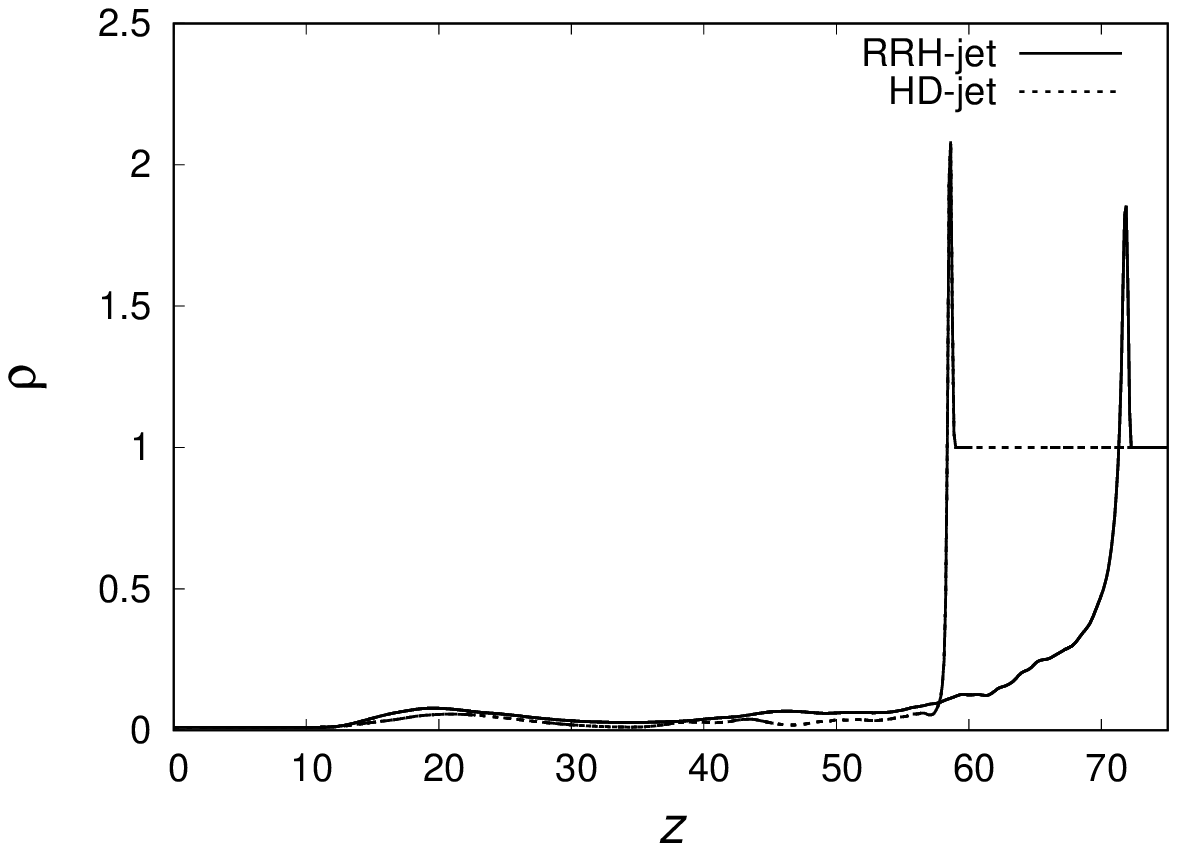}
\includegraphics[width=0.4\textwidth]{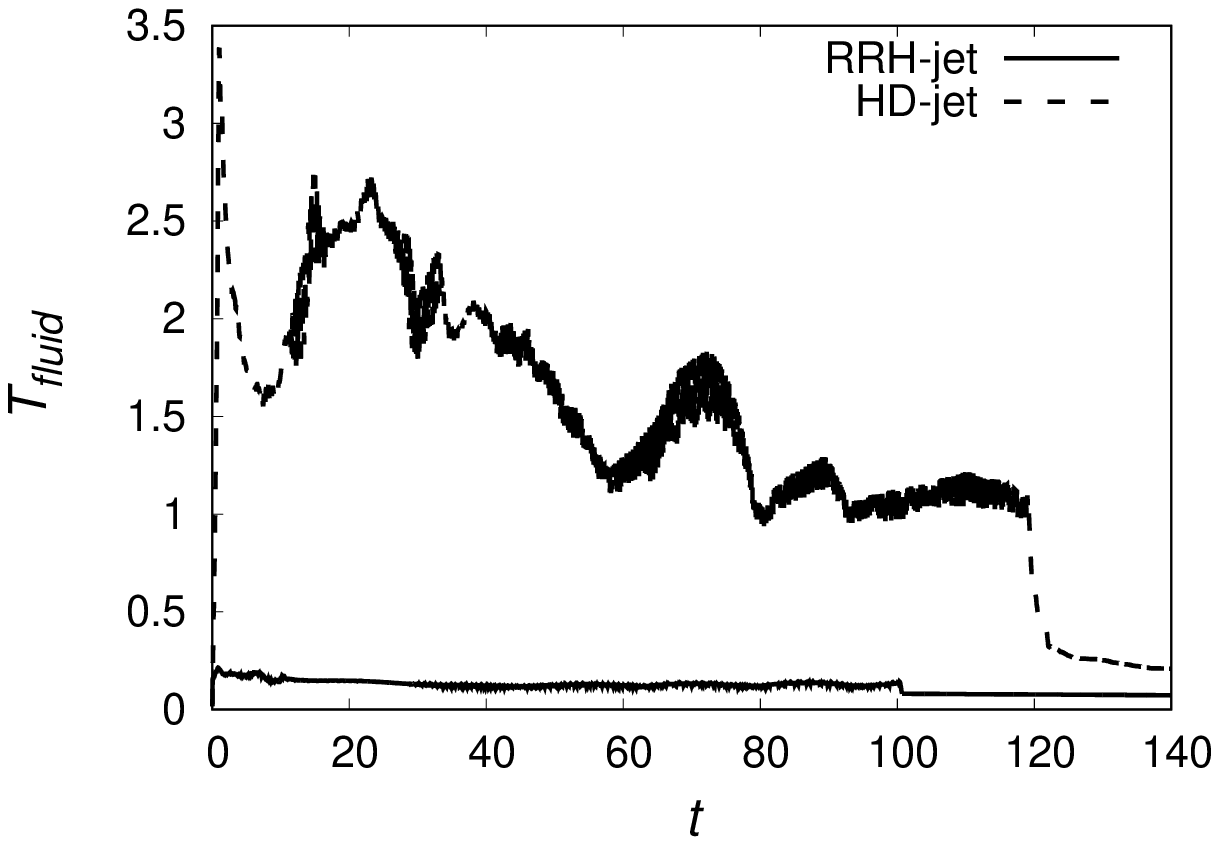}
\caption{\label{fig:comparation2} Rest-mass density profiles at $t = 100$ for the two jets, showing the  RRH-jet moves faster. The maximum of the fluid temperature for the two jets indicating the RRH-jet has a lower temperature than the fluid during the evolution.}
\end{figure}

To finalize this section, in Fig. \ref{fig:comparation2}, we show the rest-mass density profile at $t=100$ and the maximum of the fluid temperature of the jets.  We can see that the rest-mass density of the purely hydrodynamical jet is a little bigger than that of model dominated by the radiation pressure. We also find that the maximum of $T_\text{fluid}$, which is located behind the bow shock, for the HD-jet is bigger than for the RRH-jet. This agrees with Eq. (\ref{eq:tem}), from which it is expected that in regions with large optical depth, $T_\text{fluid}$ is  bigger. Then the regions where the matter is strongly coupled with  radiation, that is, with big optical depths are cooler than  regions where the matter and radiation are weakly coupled. 

\section{Final comments}
\label{sec:final_comments}

In this paper we have presented the code {\it CAFE-R},  that solves in the gray-body approximation, the relativistic Euler equations coupled with the two first moments of the Boltzmann equation for photon transport using both the Eddington and M1-closure relations.  {\it CAFE-R} has been tested over a standard ranges of optical depths and energy  in one, two, and three dimensions.  The solutions obtained with the code are consistent with  those presented in \cite{Farris_et_al_2008,Zanotti_et_al_2011,Fragile_et_al_2012,takahashi_Ohsuga_2013,Tolstov_et_al_2015} for all the test problems. This demonstrates the standard capacity of the code in the optically thin and thick regimes.

In order to check the potential of using {\it CAFE-R} in non-standard scenarios, we  presented the simulation of a highly relativistic 3D jet, spreading through a constant medium. Specifically, we  investigated the jet's response to  helical perturbations. During the evolution time $(t\sim150)$, we found that the behavior of the flow is significantly affected by the radiation field. Specifically, the perturbation on the radiation variables triggers the helicity of the jet earlier than in the HR-jet case. We also found that the fluid and the radiation depart  from thermal equilibrium in shocked regions and that the radiative jet is cooler than its hydrodynamical counterpart. 
  
The code has the ability to deal with stiff source terms, for optically thin and optically thick scenarios. However, there are some limitations that can be overtaken in future versions of the code, which are described below.

{\it The evolution method.} As mentioned before, the source terms can become stiff depending on the opacities. We use a second order accurate  IMEX Runge-Kutta scheme, that allows us to study the evolution of astrophysical scenarios with the radiation and matter  strongly coupled without any disruption. However, when the opacity is large,  this integrator is unstable. We found that there is not a unique threshold for this unstable behavior, it basically depends on the combination of three quantities, rest-mass density,  radiative energy density, and opacity coefficient. In order to explore  larger values of these variables, we need to implement higher order IMEX methods.

{\it M1-closure relation}. This closure relation improves the Eddington approximation, which cannot handle radiative transport in the optically thin limit. We have shown that the M1-closure relation gives an accurate solution even when the radiation field presents angular anisotropies (Pulse collision and Double shadow tests). However, in general, this accuracy is limited in regions where the optical depth is small and when there are multiple radiation sources. In order to improve this fact, we need to compute a closure relation directly from the Eddington tensor by using, for instance, the so-called variable Eddington tensor formalism \cite{Gehmeyr_Mihalas_1993,Stone_et_al_1992,Hayes_Norman_2003,Jiang_et_al_2012,Ohsuga_Takahashi_2016}.

{\it The GB approximation}. Our current numerical scheme is based upon frequency integrated quantities. In general, this is a good approximation that allows us to measure monochromatic radiation temperatures and bolometric light curves like those computed in \citep{Rivera_Guzman_2018}. However, in many astrophysical scenarios it is important to analyze the frequency spectrum to infer physical properties of the system under study, which cannot be done within the GB approximation. To solve this inconvenient, we need to implement a  frequency dependent scheme, for instance, the so-called multi-group scheme, which splits the frequency domain into a finite number of multi-energy groups and the equations of radiative transfer are solved within each frequency group.

{\it MHD.} Coupling  Maxwell equations to the current radiative model represents another direction of future improvements, which would be the combination with the methods used for CAFE-MHD in \citep{Lora_et_al_2015}.
Finally, we can conclude this manuscript by saying that {\it CAFE-R}  is useful to study the behavior of  relativistic radiation hydrodynamical jets and thus a potential tool to study models of gamma ray bursts. 

\acknowledgments{Acknowledgments. This research is partly supported by the following grants CIC-UMSNH 4.9, and CONACyT 258726 (Fondo Sectorial de Investigaci\'on para la Educaci\'on). The simulations were carried out in the Big Mamma cluster at the IFM and the farm funded with grant CONACyT 106466.}

\appendix

\begin{figure}
\centering
\includegraphics[width=7cm]{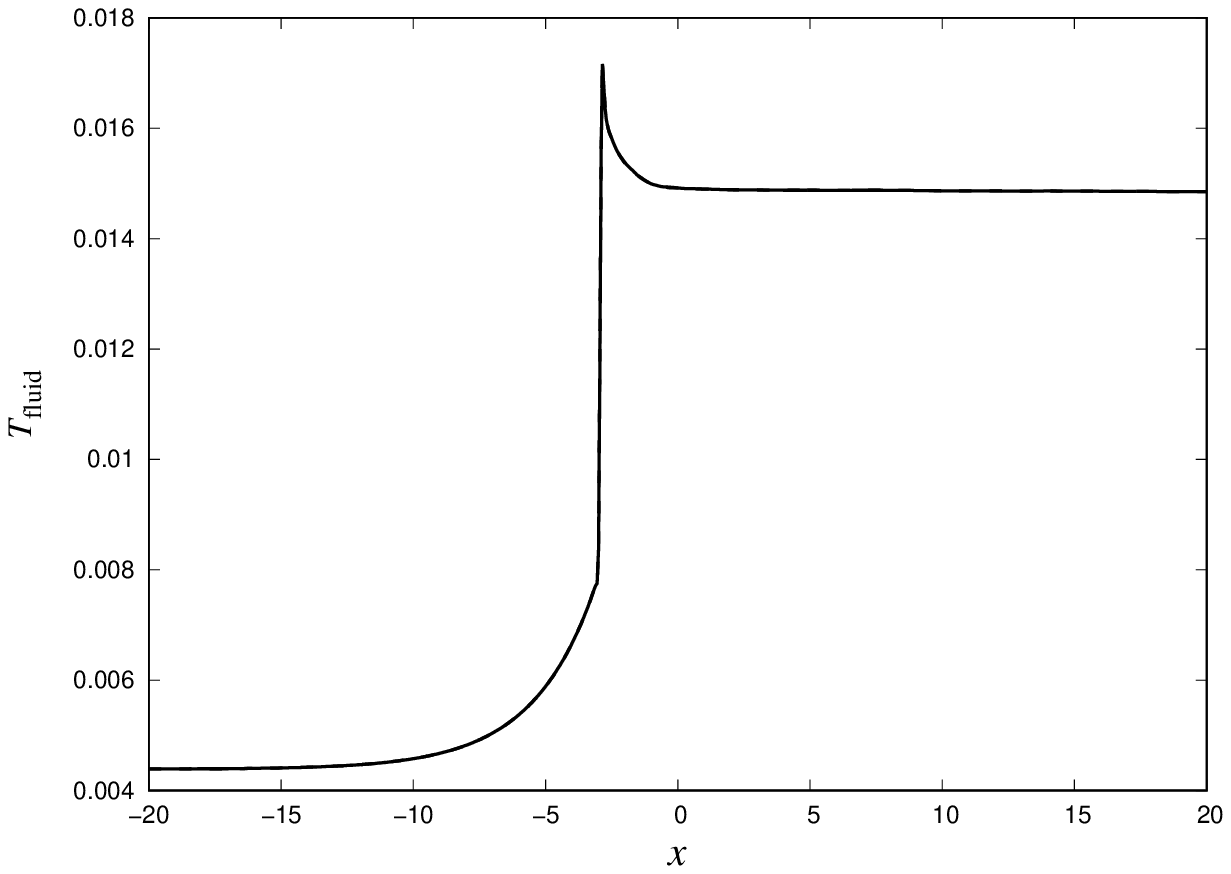}
\includegraphics[width=7cm]{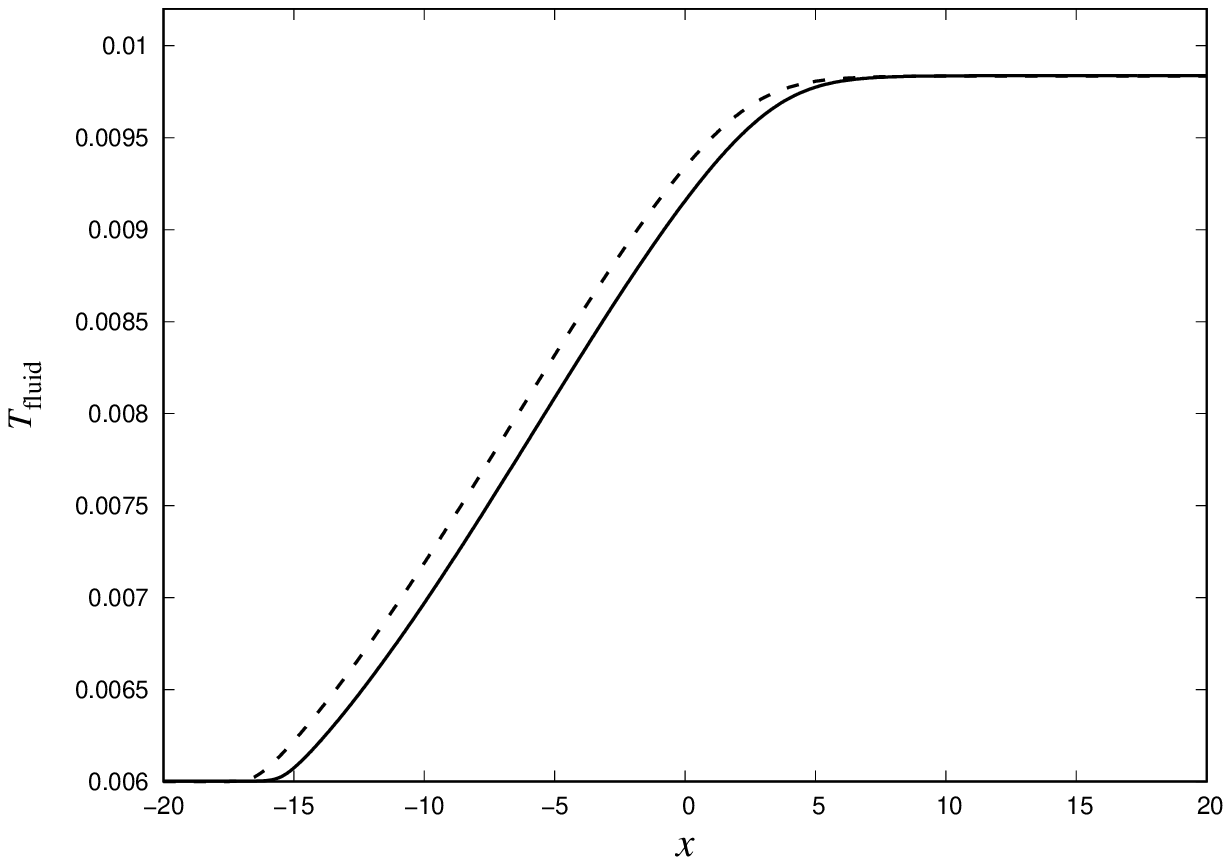}
\caption{\label{fig:appendix} (Left) Fluid temperature for Test 2 using the M1-closure at $t=3500$. (Right) Fluid temperature for Test 4a using the M1-closure at $t=3500$. Solid and dashed curves represent the fluid temperature computed from Eq.(14) and $T_{\rm fluid,\Gamma}$, respectively. Notice that for Test 2 the two curves are very similar whereas for Test 4a the differences are noticeable. }
\end{figure}

The use of Eq. (\ref{eq:tem}) is important, and it is worth to see how it influences the coupling of the fluid with radiation. As mentioned in the text, it is possible to compute the fluid temperature using $T_{\rm fluid,\Gamma}=\frac{\mu m_p}{k_B}\epsilon(\Gamma-1)$. This expression coincides  with  (\ref{eq:tem}) in the limit $P^{ij}_{\rm r}\simeq 0$. In order to illustrate the differences between the calculation of fluid temperature using $T_{\rm fluid}$ from (\ref{eq:tem}) and $T_{\rm fluid,\Gamma}$ we show in Figure \ref{fig:appendix} the results for Test 2, which is a system dominated by the fluid pressure and Test 4a, where radiation pressure dominates. The results illustrate the differences in these two regimes, so as the importance of model (14).\\

\bibliographystyle{yahapj}

\end{document}